\documentclass[12pt,epsf]{article}
\usepackage{verbatim}
\usepackage{anyfontsize}
\usepackage{dsfont}
\usepackage{tikz}
\usepackage{sidecap}
\sidecaptionvpos{figure}{t}
\usepackage{subfigure}

\usepackage[english]{babel}
\usepackage{enumitem}  

\usepackage{amssymb,amsmath}
\usepackage{graphicx, xcolor, varwidth}
\usepackage{setspace}
\usepackage[permil]{overpic}
\usepackage{cite}

\DeclareSymbolFont{matha}{OML}{txmi}{m}{it}
\DeclareMathSymbol{v}{\mathord}{matha}{118}

\colorlet{darkblue}{blue!70!black}
\colorlet{darkgreen}{green!70!black}

\usepackage[colorlinks=true,urlcolor=darkblue,linktocpage=true,linkcolor=darkblue,citecolor=darkblue]{hyperref}

\numberwithin{equation}{section}
\DeclareMathSymbol{v}{\mathord}{matha}{118}
\newcommand{\be}{\begin{equation}}
\newcommand{\ee}{\end{equation}}
\newcommand{\bea}{\begin{eqnarray}}
\newcommand{\eea}{\end{eqnarray}}
\newcommand{\bear}{\begin{eqnarray}}
\newcommand{\eear}{\end{eqnarray}}
\newcommand{\beas}{\begin{eqnarray*}}
\newcommand{\p}{\partial}
\newcommand{\eeas}{\end{eqnarray*}}
\newcommand{\ba}{\begin{array}}
\newcommand{\ea}{\end{array}}

\newcommand{\n}{\nu}

\newcommand{\pd}[2][1]{\ifnum#1=1 \frac{\partial}{\partial {#2}} \else
  \frac{\partial^#1}{\partial {#2}^{#1}}\fi}
\newcommand{\dpd}[2][1]{\ifnum#1=1 \dfrac{\partial}{\partial {#2}} \else
  \frac{\partial^#1}{\partial {#2}^{#1}}\fi}
\newcommand{\td}[2][1]{\ifnum#1=1 \frac{d}{d{#2}} \else
  \frac{d^#1}{d{#2}^{#1}}\fi}

\newcommand{\x}{\xi}

\newcommand{\nbox}{{\,\lower0.9pt\vbox{\hrule \hbox{\vrule height 0.2 cm \hskip 0.19 cm \vrule height 0.2 cm}\hrule}\,}}

\newcommand{\ie}{{\it i.e.,}\ }

\def\O{{\cal O}}

\newcommand{\half}{\tfrac{1}{2}}

\newcommand{\bz}{\bar{z}}



\begin{document}
\begin{spacing}{1.3}
\begin{titlepage}

\begin{center}
{\Large \bf 
Shockwaves from the Operator Product Expansion
}

\vspace*{6mm}

Nima Afkhami-Jeddi,$^*$ Thomas Hartman,$^*$ Sandipan Kundu,$^{*\dagger}$ and Amirhossein Tajdini$^*$

\vspace*{6mm}

\textit{$^*$Department of Physics, Cornell University, Ithaca, New York, USA\\}

\vspace{3mm}

\textit{$^\dagger$Department of Physics and Astronomy, Johns Hopkins University,
Baltimore, Maryland, USA\\}

\vspace{6mm}

{\tt \small na382@cornell.edu, hartman@cornell.edu, kundu@cornell.edu, at734@cornell.edu}

\vspace*{6mm}
\end{center}
\begin{abstract}

We clarify and further explore the CFT dual of shockwave geometries in Anti-de Sitter. The shockwave is dual to a CFT state produced by a heavy local operator inserted at a complex point. It can also be created by light operators, smeared over complex positions. We describe the dictionary in both cases, and compare to various calculations, old and new. In CFT, we analyze the operator product expansion in the Regge limit, and find that the leading contribution is exactly the shockwave operator, $\int du h_{uu}$, localized on a bulk geodesic. For heavy sources this is a simple consequence of conformal invariance, but for light operators it involves a smearing procedure that projects out certain double-trace contributions to the OPE. We revisit causality constraints in large-$N$ CFT from this perspective, and show that the chaos bound in CFT coincides with a bulk condition proposed by Engelhardt and Fischetti. In particular states, this reproduces known constraints on CFT 3-point couplings, and confirms some assumptions about double-trace operators made in previous work.

\end{abstract}

\end{titlepage}
\end{spacing}

\vskip 1cm
\setcounter{tocdepth}{2}  
\tableofcontents

\begin{spacing}{1.3}

\section{Introduction}

\begin{figure}[p]
\begin{center}
\includegraphics[width=3in]{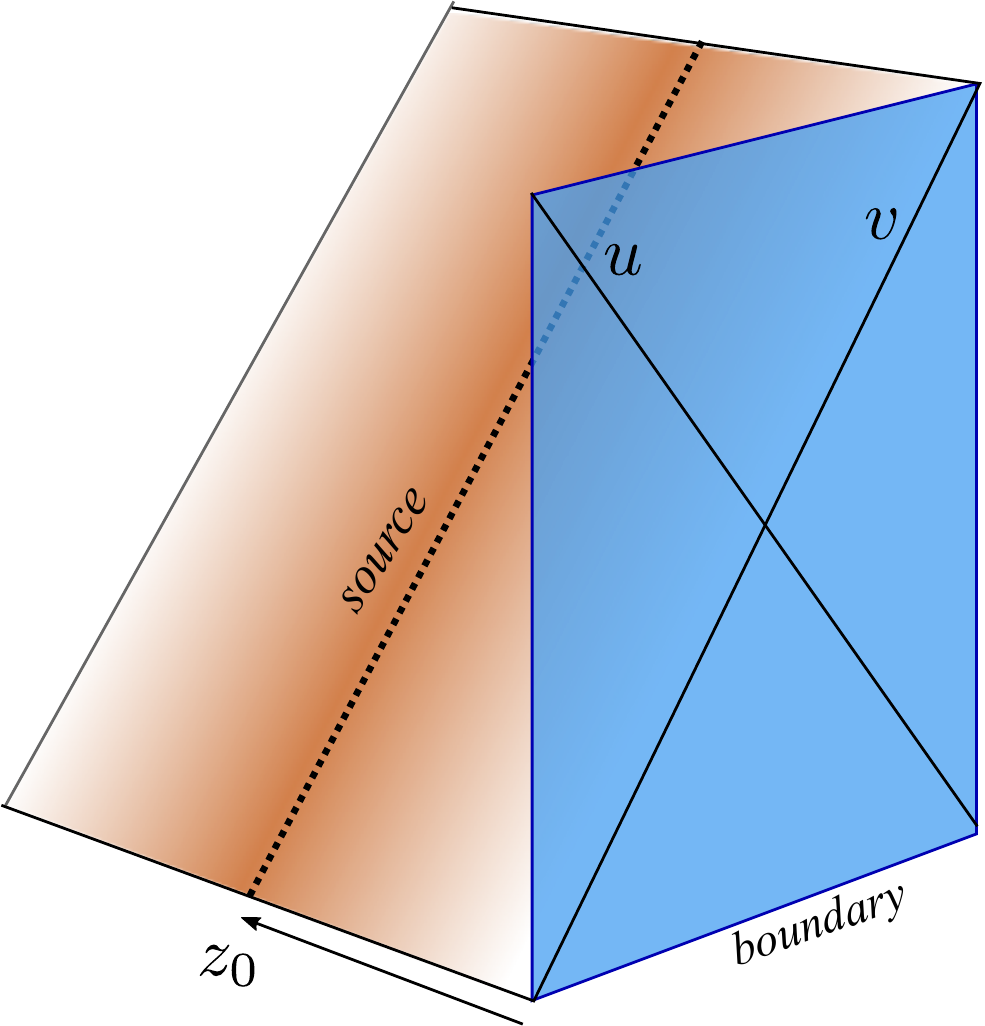}
\end{center}
\caption{
Planar shockwave \eqref{swmetric} in Poincare coordinates. The source travels on a null geodesic parallel to the boundary, at fixed radial coordinate $z = z_0$. The shockwave is on the null plane $u=0$. 
\label{fig:planarshock}}
\end{figure}

A shockwave geometry describes the gravitational field of a point source traveling on a null geodesic \cite{Aichelburg:1970dh}. In Anti-de Sitter \cite{Hotta:1992qy,Horowitz:1999gf}, it plays a key role in the study of various phenomena in gravity and conformal field theory \cite{Cornalba:2006xk,Cornalba:2006xm,Cornalba:2007zb,Hofman:2008ar,Hofman:2009ug,Nozaki:2013wia,Shenker:2013pqa,Asplund:2014coa,Camanho:2014apa, Kulaxizi:2017ixa, Costa:2017twz,causality1,causality2,Hartman:2016lgu,Afkhami-Jeddi:2016ntf}. One incarnation of this solution has the source traveling parallel to the boundary, at fixed radial distance $z=z_0$ in Poincare coordinates, as shown in figure \ref{fig:planarshock}. The resulting gravitational shock meets the boundary along a null plane, while the source itself hits the boundary only at null infinity.

In the dual CFT, this state can be created by inserting a momentum-space operator smeared against a wavepacket profile \cite{Cornalba:2006xk,Cornalba:2006xm,Cornalba:2007zb}. It can also be created by inserting a heavy local operator $\psi$, with dimension $\Delta_\psi \gg 1$, at a single point \cite{Nozaki:2013wia,Roberts:2014isa,Asplund:2014coa,causality1}. Yet another proposal to create a shockwave with a different, complexified wavepacket was described in \cite{ Afkhami-Jeddi:2016ntf}.  These three constructions differ in their microscopic details, but should produce all of the same observables for probes that avoid the delta function at the shockwave source. 

In this paper we will work out various aspects of the dictionary relating (linearized) AdS shockwaves to CFT operator insertions, focusing on the latter two constructions:  the heavy local operator, and the complexified wavepacket. Parts of these results have been used implicitly in previous work \cite{Nozaki:2013wia,Roberts:2014isa,Caputa:2014vaa,causality1}.

\begin{figure}[p]
\begin{center}
\includegraphics[width=3in]{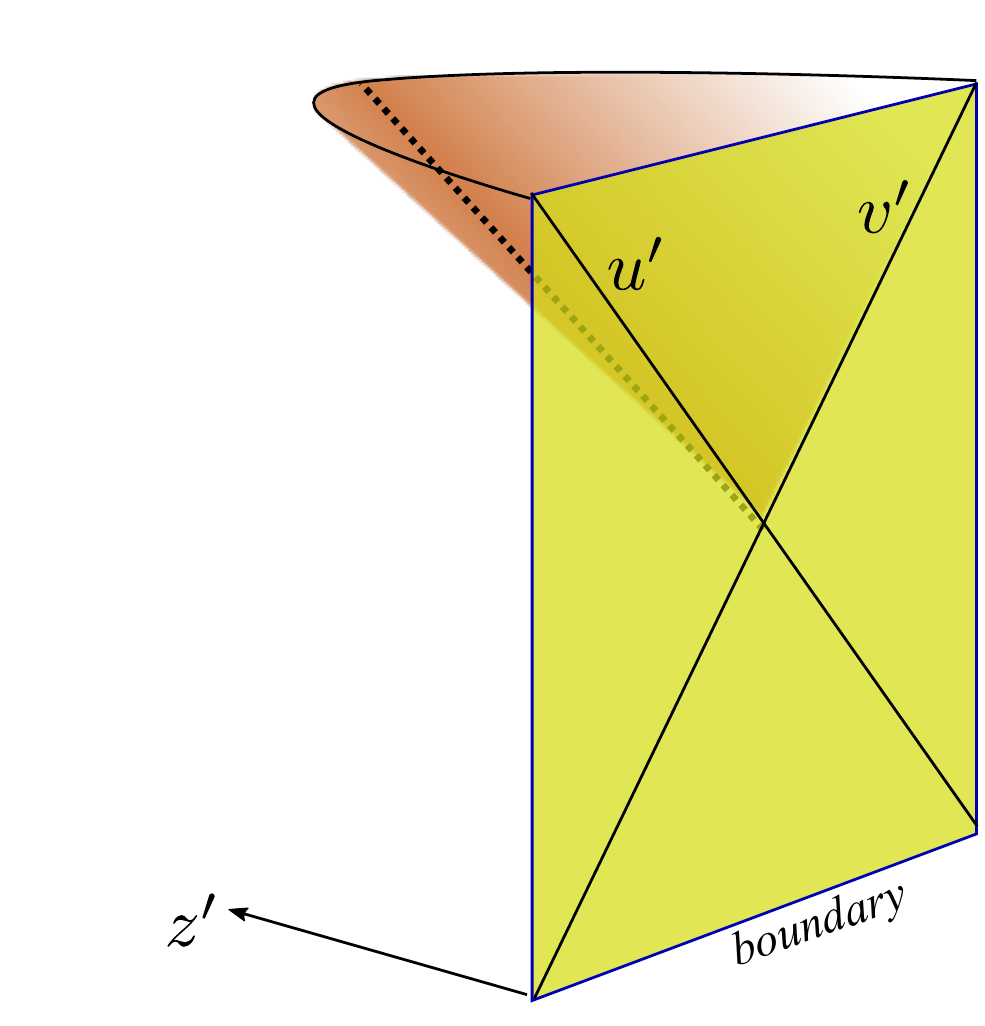}
\end{center}
\caption{
Spherical shockwave \eqref{sw_HI} in Poincare coordinates. The source travels on a radial null geodesic that hits the boundary at the origin, $u' = v'= 0$. The resulting shockwave is a null cone.
\label{fig:sphericalshock}}
\end{figure}

Our main new result is a relationship between shockwaves in the bulk, and the operator product expansion (OPE) in CFT.  The propagation of a probe field $\phi$ on the $\psi$-shockwave  computes a four-point function, $\langle \psi \phi\phi  \psi\rangle$, in a highly boosted kinematics dubbed the Regge limit \cite{Brower:2006ea,Cornalba:2006xk,Cornalba:2006xm,Cornalba:2007zb,Cornalba:2007fs,Cornalba:2008qf,Costa:2012cb}. It is well known that the crossing equation for this four-point function leads to a rich story relating double-trace anomalous dimensions in CFT to the graviton propagator in the bulk \cite{Cornalba:2006xk,Cornalba:2006xm,Cornalba:2007zb,Heemskerk:2009pn,Komargodski:2012ek,Fitzpatrick:2012yx,Alday:2016htq,Camanho:2014apa, Kulaxizi:2017ixa, Costa:2017twz}. We will essentially strip off the probes $\phi\phi$ from this story, and directly examine the $\psi\psi$ OPE in the Regge limit.  The leading term is precisely the bulk shockwave operator.

Of course, the correlators and the OPE ultimately contain the same information, but the OPE point of view is useful for several reasons.  First, analogous results in the lightcone limit were a crucial step toward deriving the averaged null energy condition \cite{Hartman:2016lgu} and quantum null energy condition \cite{Balakrishnan:2017bjg} from causality. Second, it makes  clear why various bulk and boundary results must match, without actually doing the calculations, and extends this match --- including causality constraints --- to a more general class of quantum states. Finally, it generalizes existing constraints to higher-point functions and multiple shocks, though we will not explore this in detail here.

In addition to the Regge OPE, our other main result is a technical improvement on our previous work on large-$N$ causality constraints \cite{ Afkhami-Jeddi:2016ntf}.  We use conformal Regge theory to confirm an assumption made there regarding the contributions of double-trace operators to smeared, spinning conformal correlators, in the shockwave regime. This is a natural extension of the Regge OPE formula to spinning operators.

In the rest of this introduction we give a brief summary of the main results.

\subsection{Shockwaves from heavy operators}

 The simplest way to create a shockwave is by inserting a heavy scalar primary operator $\psi$, since this does not require any smearing.  It does, however, require somewhat exotic kinematics, with operators inserted near $i\infty$. For example, the shockwave depicted in figure \ref{fig:planarshock}, with a real metric (see \eqref{swmetric}), is dual to the CFT state
\be\label{introdict}
|\Psi\rangle  =  \psi(u = \frac{i\Delta_\psi}{2E}, v = \frac{2iz_0^2 E}{\Delta_\psi} )|0\rangle  \ ,
\ee
where $(u,v)$ are null coordinates in Minkowski, $E$ is the the energy of the shockwave, and the duality holds for $1 \ll \Delta_\psi \ll E$. Analogous results apply to the spherical shockwave in figure \ref{fig:sphericalshock}, which is created by a local operator inserted near the origin, and is related to the planar shockwave by a null inversion.
We start in section \ref{s:dictionary} by deriving this dictionary and comparing the stress tensor one-point functions to CFT.

Next, in section \ref{s:reggeope}, we turn to the relationship between shockwaves and the OPE. In the lightcone limit, the leading term in the OPE is the averaged null energy operator $\int  du T_{uu}$ \cite{Hartman:2016lgu}.  In the Regge limit, we will show that the leading contribution is
\be\label{introres}
\psi(-u,-v)\psi(u,v) \  \propto  \ 1 - i E z_0^2 \int_{\gamma}du h_{uu}
\ee
as $u\to \infty$ with $uv$ fixed. The ``shockwave operator" on the right-hand side is the bulk metric perturbation integrated over a null geodesic through AdS$_{d+1}$, with $d \geq 3$. It can also be viewed as the linearized length operator. Physically, it corresponds to the source for a planar shockwave, or to the time delay in a background produced by other operators.

Although \eqref{introres} looks like a bulk formula, it can be interpreted as a CFT operator by translating $h_{uu}$ into CFT language. In the lightcone limit, $h_{uu} \sim T_{uu}$, so this reproduces the averaged null energy operator derived in \cite{Hartman:2016lgu}. More generally, the translation to CFT follows from the HKLL formula \cite{Kabat:2012hp} for $h_{uu}$. Interpreted this way, and keeping only the leading single-trace term, \eqref{introres} gives a formula for the stress tensor contribution in the Regge limit, in any CFT. In CFT language, this operator is the part of the OPE block \cite{Czech:2016xec,deBoer:2016pqk} that grows in the Regge limit.  In a holographic theory, \eqref{introres} contains more information than the OPE block, since it also accounts for multitrace contributions.  

\subsection{Shockwaves from light operators}

The above discussion applies only to operators with $\Delta_\psi \gg 1$.  For a light operator $O$ --- and in particular, for the stress tensor $T_{\mu\nu}$ which will be important for the discussion of causality --- the OPE is not given by \eqref{introres}. In gravity language, a light operator inserted at a point creates a wave that spreads, not a particle that travels on a geodesic. To remedy this, we insert a wavepacket. The smeared operators behave just like heavy operators, in that they create a state dual to the bulk shockwave, and have exactly the same Regge OPE \eqref{introres}, with $(u,v)$ now interpreted as the wavepacket centers. Importantly, this OPE has no contributions from the $O^2$ double-trace operators that would appear in the unsmeared OPE.  We will derive this statement from the bulk by smearing a Witten diagram vertex, and from CFT using conformal Regge theory \cite{Costa:2012cb}. It applies to both scalar and spinning operators $O$. 

In the scalar case, the smeared operators $\tilde{O}$ have exactly the same Regge OPE as the heavy operator $\psi$ in \eqref{introres}. For  spinning operators, rather than deriving an OPE formula like \eqref{introres} explicitly, we work directly with the four-point function, and show that smearing projects out the double-trace contributions to the conformal Regge amplitude. The procedure is based closely on \cite{Kulaxizi:2017ixa, Costa:2017twz}, and differs only by the fact that our wavepackets are rotated into a complex direction, following \cite{Afkhami-Jeddi:2016ntf}. This choice is motivated by the chaos bound \cite{Maldacena:2015waa}, which applies only in a kinematic regime where certain points are spacelike separated, and cannot be applied to ordinary wavepackets smeared over real spacetime points. (Other arguments, rather than the chaos bound, were used to derive causality constraints from real wavepackets in \cite{Kulaxizi:2017ixa, Costa:2017twz}; see also \cite{Caron-Huot:2017vep}.)

Exactly the same complex smearing procedure (up to a conformal transformation) was used in our previous work on large-$N$ causality constraints \cite{Afkhami-Jeddi:2016ntf}.  There, we suggested on other grounds that the double-trace contributions should drop out, but did not show it explicitly; the present result confirms this, and therefore closes this potential loophole in the argument.

\subsection{Causality constraints}

In a CFT with a gravity-like holographic dual, the shockwave operator in \eqref{introres} (including multitraces) dominates the 4-point function, so it is subject to the causality constraints on conformal correlators derived in \cite{Maldacena:2015waa,Afkhami-Jeddi:2016ntf}. Therefore $\int_{\gamma} du h_{uu}$ has a positive expectation value in states defined perturbatively about the vacuum. This generalizes the averaged null energy condition (ANEC) $\int du T_{uu} \geq 0$, and reduces to the ANEC in a lightcone limit. 

The ANEC has also been derived from the monotonicity of relative entropy \cite{Balakrishnan:2017bjg}. It would be very interesting to understand the Regge analogue of that calculation, or more generally, the constraints from quantum information in the Regge limit.

If the state has a geometric dual, then the sign constraint $\int_{\gamma} du h_{uu} \geq 0$ coincides with the bulk causality condition of Engelhardt and Fischetti \cite{Engelhardt:2016aoo}. Noteably, it is apparently weaker than the averaged null curvature condition $\int du R_{uu} \geq 0$, or bulk null energy condition, which were the starting point for Gao and Wald \cite{Gao:2000ga} to prove that asymptotically AdS spacetimes satisfy boundary causality. Our constraint also applies to non-geometric states, such as superpositions of different geometries.

We postpone a detailed discussion of causality constraints to section \ref{s:causality}, but the brief version is that this inequality encompasses, and generalizes, a number of known causality constraints in CFT: the scalar causality constraints derived in \cite{causality1}, the Hofman-Maldacena bounds \cite{Hofman:2008ar, Hofman:2009ug, causality2,Hofman:2016awc}, the averaged null energy condition \cite{Kelly:2014mra,Faulkner:2016mzt,Hartman:2016lgu}, and the $a=c$ type bounds derived on the gravity side in \cite{Camanho:2014apa} and the CFT side in \cite{Afkhami-Jeddi:2016ntf,Kulaxizi:2017ixa,Costa:2017twz}. In holographic CFTs, the result here is more general, since we will show that the operator $\int_{\gamma} du h_{uu}$ is positive in a wider class of states. It can be evaluated in particular states to reproduce each of these previous constraints.

\begin{figure}[p]
\begin{center}
\includegraphics[width=2.5in]{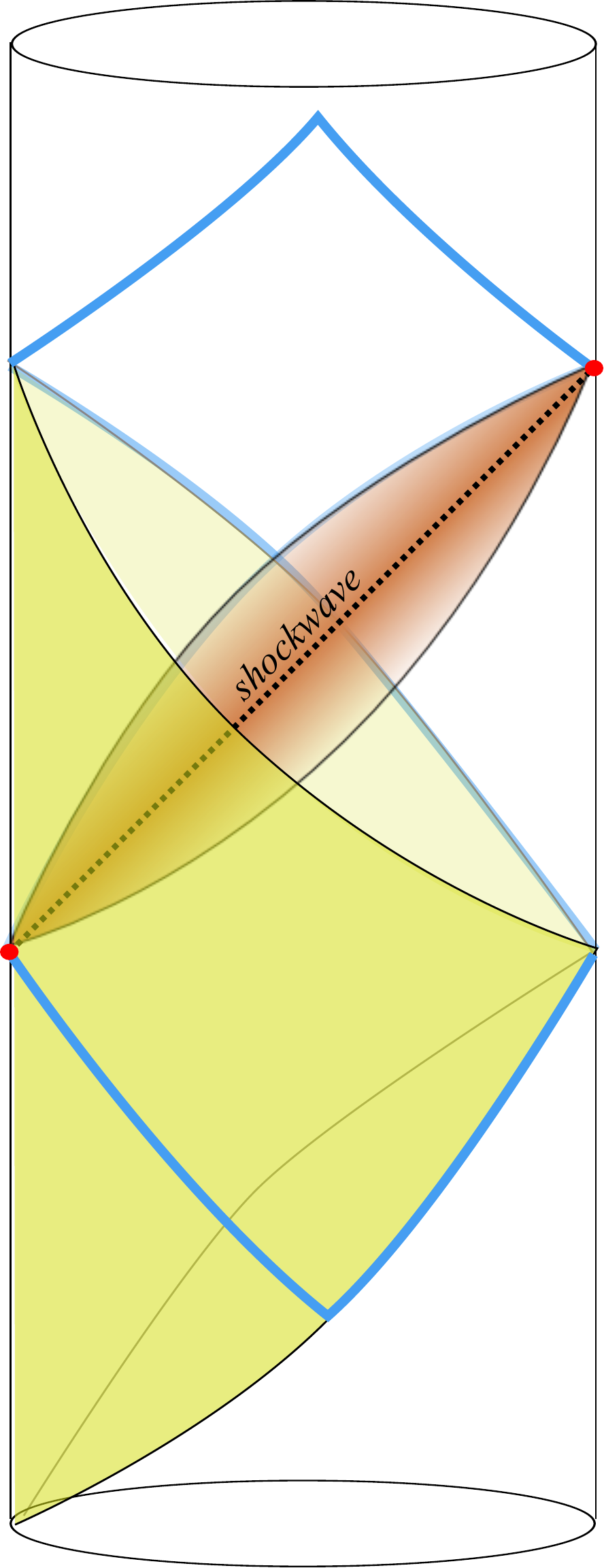}
\end{center}
\caption{
Shockwave in global AdS, where the boundary is the Lorentzian cylinder. The dashed line is the source, which hits the boundary at the red dots. This source creates a shockwave on the shaded null surface. In the shaded-yellow Poincare patch, the solution is the spherical shockwave. In the shifted Poincare patch, shown as a thick blue outline, the solution is the planar shockwave. The yellow and blue patches in this figure correspond to the same color-coding as the other figures.
\label{fig:cylinder}}
\end{figure}

\section{Shockwaves from Heavy Operators}\label{s:dictionary}

In this section we review the shockwave metric and the CFT construction using a heavy local operator insertion. Most of these results are known in some form, but we will start from the beginning and clarify a few points along the way. The main goal is to derive the dictionary \eqref{introdict} for the planar shock.  This result holds in general dimensions, $AdS_{d+1}$ with $d \geq 2$.

To summarize briefly, it is easiest to start with the dictionary for the spherical shockwave \cite{Hotta:1992qy,Horowitz:1999gf}, depicted in figure \ref{fig:sphericalshock}. In this geometry, the source particle hits the boundary at the origin, $u'=v'=0$, so it is roughly  dual to the CFT state with an operator inserted there, $\psi(0)|0\rangle$. `Roughly' because such a state is not normalizable, so it is useful to regulate it by moving the source slightly into Euclidean time, inserting the operator at $t = i\delta$. In the bulk, this corresponds to a geometry where the source does not quite hit the boundary, but has closest approach $z = \delta$. In the limit $\delta \to 0$, it becomes the shockwave geometry.

The planar shockwave (figure \ref{fig:planarshock}) and the spherical shockwave (figure \ref{fig:sphericalshock}) are related by a null inversion. This is a conformal transformation that sends $v' \to -1/v$. The effect is easiest to understand on the Minkowski cylinder; see figure \ref{fig:cylinder}. The original $(u',v')$ coordinates cover a diamond on the cylinder. The null inversion sends $v' =0 \to v= -\infty$, so the new $(u,v)$ coordinates cover a new diamond, shifted in the null direction by `half of a patch'.

Since the original insertion was at $v=0$, the operator insertion for a planar shockwave should be roughly at $-\infty$. Carefully keeping track of the Euclidean-time regulator gives the precise statement of the dictionary, \eqref{introdict}.

We will derive these results in the opposite order, starting with the planar shockwave, then performing the null inversion to reproduce the spherical shockwave.

\subsection{Shockwave solutions in AdS}

\subsubsection{Planar coordinates}
The metric of the planar shockwave in AdS$_{d+1}$ \cite{Hotta:1992qy,Horowitz:1999gf} is
\be\label{swmetric}
ds^2= 
\frac{L^2}{z^2}(-du dv +  dz^2+d \vec{x}^2) + h_{uu}^{Shock}du^2
\ee
with 
\begin{align}
h_{uu}^{Shock} & = \notag\\
& E \frac{16\pi G_N  (4\pi) ^{\frac{1-d}{2}}\Gamma \left(\frac{d+1}{2}\right)z_0}{z L^{d-1}d(d-1) }  \left(\frac{\rho ^2}{1-\rho ^2}\right)^{1-d}  \, _2F_1\left(d-1,\frac{d+1}{2};d+1;1 - \frac{1}{\rho^2}\right)\delta(u)\ .\label{huu}
\end{align}
We will assume that the source is localized at $u=0, z = z_0, \vec{x}=0$ and hence $\rho$ is given by \cite{Camanho:2014apa}
\be
\rho=\sqrt{\frac{(z-z_0)^2+\vec{x}^2}{(z+z_0)^2+\vec{x}^2}}\ .
\ee
This geometry is illustrated in figure \ref{fig:planarshock}. The source travels on a null geodesic at fixed radial distance, $z = z_0$, and the geometry solves the Einstein equation with this source, which reduces to
\be
z^{d-1}\p_z\left[z^{1-d}\p_z(z^2 h_{uu}^{Shock})\right] + z^2 \vec{\p}^{\,2} h_{uu}^{Shock} = -16\pi G_N E z_0^{d-1}\delta(u)\delta^{d-2}(\vec{x})\delta(z-z_0) \ .
\ee

\subsubsection{Holographic stress tensor}
According to the usual holographic dictionary, the boundary stress tensor is proportional to the $O(z^{d-2})$ components of the metric \cite{dhss,skenderis,skenderis2}.
 For the planar shockwave \eqref{swmetric}, this gives
\be\label{gravityT}
\left\langle T_{uu}(u,v,\vec{x})\right\rangle=E \frac{2^{d-2}  \ \Gamma \left(\frac{d-1}{2}\right) z_0^{d} }{ \pi ^{\frac{d-1}{2}}(z_0^2+\vec{x}^2)^{d-1}} \delta(u) \ .
\ee
Other components vanish. Integrating gives the total energy, $E$.

\subsubsection{Spherical coordinates}

The spherical shockwave can be obtained from \eqref{swmetric} by a coordinate change \cite{Hotta:1992qy,Horowitz:1999gf}:
\be\label{fourshift1}
u=u'-\frac{\vec{x}'^2}{v'}-\frac{z'^2}{v'}\ ,
\quad v=-\frac{z_0^2}{v'}\ , 
\quad z=\frac{z' z_0}{v'}\ , 
\quad x_i=\frac{x'_i z_0}{v'}\ . 
\ee
This acts as a conformal transformation on the boundary which we refer to as the null inversion. If we write $u'=t'-y'$, $v'=t'+y'$, and $r^2=y'^2+\vec{x}'^2$, then under this transformation, the metric  (\ref{swmetric}) becomes 
\be\label{sw_HI}
ds^2=\frac{L^2}{z'^2}\left(dz'^2-dt'^2+dr^2+r^2 d\Omega^2\right)+ds^2_p\ ,
\ee
where the shockwave part, coming from $h_{uu}$ in \eqref{swmetric}, is 
\be
ds^2_p=\frac{L^2}{z'^2}\left(\frac{2tz'}{z_0}\right)f\left(\frac{t'-z'}{z'}\right)\delta\left(t'-\sqrt{r^2+z'^2}\right)d\left(t'-\sqrt{r^2+z'^2}\right)^2
\ee
with
\begin{align}
f\left(\frac{t'-z'}{z'}\right) =\frac{16\pi G_N E \pi ^{\frac{1-d}{2}}\Gamma \left(\frac{d+1}{2}\right)z_0}{L^{d-3}(d-1) d} \left(\frac{t'-z'}{z'}\right)^{1-d}  \, _2F_1\left(d-1,\frac{d+1}{2};d+1;\frac{2z'}{z'-t'}\right)\ .
\end{align}
This solution is shown in figure \ref{fig:sphericalshock}. The source particle travels radially, on a null geodesic that hits the boundary at the origin, and the resulting shockwave is an expanding spherical shell. The metric \eqref{sw_HI} can also be obtained from an infinitely boosted AdS-Schwarzschild black hole, where the mass of the black hole is scaled down to hold the total energy fixed.\footnote{The coordinate change for comparison to \cite{Horowitz:1999gf} is
\be
y_+=\frac{t'^2-z'^2-r^2}{z'+t'}\ , \qquad y_-=-\frac{L^2}{z'+t'}\ , \qquad \sum_{i=2}^dy_i^2=\frac{r L}{z'+t'}\ .
\ee
} 

The boundary stress tensor obtained from \eqref{sw_HI} is supported on the null cone,
\be\label{sthi}
\langle T_{\bar{u}\bar{u}}\rangle= E \frac{\pi ^{\frac{1}{2}-\frac{d}{2}} ~\Gamma \left(\frac{d-1}{2}\right)}{ r^{d-2}}\delta(\bar{u})\ ,
\ee
where $\bar{u}=t'-r$. This shockwave has total energy $2E$, twice that of the planar shockwave, because part of the expanding spherical shell lies at null infinity in the planar coordinates and was not included there.

\subsubsection{Global picture}
The important thing to notice in the coordinate change \eqref{fourshift1} is that it involves the null inversion $v \sim -1/v'$. This maps the origin in the primed coordinates to $I^-$ in the unprimed coordinates.  This makes sense, since the source particle hits the boundary at the origin of the spherical shock, but only at null infinity in the planar shock. The shockwave solution in global coordinates is illustrated in figure \ref{fig:cylinder}. The $(u,v)$ and $(u',v')$ coordinates cover two different Poincare patches, related by a null shift.

\subsection{Shockwave states in CFT}\label{SW_CFT}

Now we will describe how to create these states in CFT. As mentioned in the introduction, there are a number of different ways to do this, which all produce the same physics away from the point-particle source. There is also the choice of whether we want to reproduce expectation values in the `in-in' or `in-out' sense. We will start with the simplest and most intuitive case: in-in correlators in the spherical shockwave created by a heavy local operator. This state has also been discussed in  \cite{Nozaki:2013wia,Asplund:2014coa,causality1,causality2}. Much of this applies to any CFT, whether or not it is holographic, so for now the discussion is general and we will specialize to holographic CFTs later. We restrict to $d>2$.

\subsubsection{Insertion at the origin}\label{origin}

A pure state in CFT can be created by inserting a local operator, $\psi(x)|0\rangle$. (We will always take space in the CFT to be $R^{d-1}$, so the spectrum is continuous.) If $x$ is a point in the Lorentzian spacetime, this state is not normalizable, but an operator insertion at non-zero Euclidean time produces a normalizable state.  We first consider a shockwave state created by inserting $\psi$ at $y'=0$, $t'=i\delta$ and $\vec{x}'=0$, with $\delta$ small. In null coordinates $v'=t'+y'$ and $u'=t'-y'$, with the metric $ds^2=-du'dv'+d\vec{x}'^2$, we have
\be\label{psiz}
|\psi_0\rangle=\psi(x_0)|0\rangle\ , \qquad \langle\psi_0|=\langle 0| \psi(x_0^*)\ .
\ee
where
\be\label{centerpoints}
x_0=(u_0',v_0',\vec{x}_0')=(i\delta,i\delta,\vec{0})\ , \qquad x_0^{*}=(-i\delta,-i\delta,\vec{0})\ .
\ee
The subscript $0$ is a reminder that $\psi$ is inserted near the origin, since another choice is considered below. The conjugate state in \eqref{psiz} is defined with respect to the usual Hermitian conjugate in Minkowski space, which does nothing to real operators inserted at real Minkowski points, but conjugates the coordinates when they are complex, in particular reflecting Euclidean time $i t \to - i t$.

Now let's compute the stress tensor expectation value in the state $|\psi_0\rangle$. The three-point function of the stress tensor and two scalars, in $d$ spacetime dimensions, is entirely fixed by conformal invariance \cite{Osborn:1993cr}, 
\be\label{OP}
\langle T_{\mu\nu}(x_1) \psi(x_2) \psi(x_3)\rangle=\frac{a_d}{x_{12}^d x_{23}^{2\Delta_\psi-d}x_{31}^d}\left(\frac{X_\mu X_\nu}{X^2}-\frac{\eta_{\mu \nu}}{d} \right)\ ,
\ee
where
\be
x_{IJ}=|x_I-x_J|\ , \qquad X^\mu=\frac{x_{13}^\mu}{x_{13}^2}-\frac{x_{12}^\mu}{x_{12}^2}\ , \qquad X^2=\frac{x_{23}^2}{x_{13}^2x_{12}^2}
\ee
and the Ward identity fixes
\be
a_d = - \Delta_\psi \frac{ \Gamma(d/2) d}{2\pi^{d/2}(d-1)}\ . \label{ad}
\ee
We have normalized the scalar by $\langle \psi(x_1) \psi(x_2) \rangle = |x_1-x_2|^{-2\Delta_\psi}$, and the stress tensor is canonically normalized (for example by the Noether procedure).

This can be used to calculate the expectation value
\be\label{texp}
\langle T_{\mu\nu}(x)\rangle=\frac{\langle\psi_0| T_{\mu\nu}(x)| \psi_0 \rangle}{\langle \psi_0|\psi_0\rangle} =\frac{ \langle \psi(x_0^*) T_{\mu\nu}(u, v, \vec{x}) \psi(x_0) \rangle }{\langle \psi(x_0^*)  \psi(x_0) \rangle}\ .
\ee
Using \eqref{OP}, this formula, at finite $\delta$, agrees exactly with the boundary stress tensor produced by a boosted AdS-Schwarzschild black hole, at finite boost \cite{Horowitz:1999gf}.\footnote{Explicitly, for $\vec{x}' = 0$ (and dropping the primes),
\begin{align}
\langle T_{uu}(u, v, \vec{x}=0) \rangle&= - \frac{2^{d-2} a_d\delta ^d }{(u^2+\delta^2)^{1+\frac{d}{2}}(v^2+\delta^2)^{-1+\frac{d}{2}}}\ ,\label{st1} &
\langle T_{vv}(u, v, \vec{x}=0) \rangle&= - \frac{2^{d-2} a_d\delta ^d}{(u^2+\delta^2)^{-1+\frac{d}{2}}(v^2+\delta^2)^{1+\frac{d}{2}}}\ ,\notag\\
\langle T_{uv}(u, v, \vec{x}=0) \rangle&= -\frac{2^{d-2} a_d\delta ^d(d-2)}{d(u^2+\delta^2)^{d/2}(v^2+\delta^2)^{d/2}}\ ,\notag &
\langle T_{ij}(u, v, \vec{x}=0) \rangle &= - \frac {2^d a_d \delta ^d \delta_{ij}}{d(u^2+\delta^2)^{d/2}(v^2+\delta^2)^{d/2}}\notag\ .
\end{align}
}
To take the limit $\delta\rightarrow 0$, note that
\be\label{regdelta}
\lim_{\delta \to 0} \frac{\delta^{n+1}}{(z^2+\delta^2)^{1+\frac{n}{2}}} = \frac{\sqrt{\pi } \Gamma \left(\frac{n+1}{2}\right)}{  \Gamma \left(\frac{n}{2}+1\right)}\delta(z) \ .
\ee
Therefore
\be
\langle T_{\bar{u}\bar{u}}\rangle=\frac{ \Delta_\psi \Gamma \left(\frac{d-1}{2}\right)}{2\pi^{\frac{d-1}{2}} r^{d-2}\delta}\delta(\bar{u})\ .\label{st5}
\ee
where $\bar{u} = t'-r$, $\bar{v} = t'+r$, and $r^2 = y'^2 + \vec{x}'^2$. Thus the stress-energy is supported on a lightcone emanating from the origin. This stress tensor exactly matches with the boundary stress tensor \eqref{sthi} computed for the AdS spherical shockwave \eqref{sw_HI} once we identify
\be\label{pu}
E =\frac{ \Delta_\psi}{2  \delta}\ .
\ee
This calculation did not assume holography or any particular $\Delta_\psi$, but we will see later that in a holographic theory, this state is dual to a localized shockwave in the bulk when $c_T \gg \Delta_\psi \gg 1$, where $c_T \sim N^2$ is the coefficient in $\langle TT\rangle$.

\subsubsection{Insertion at infinity}\label{infinity}
\begin{figure}[t]
\begin{center}
\includegraphics[width=3in]{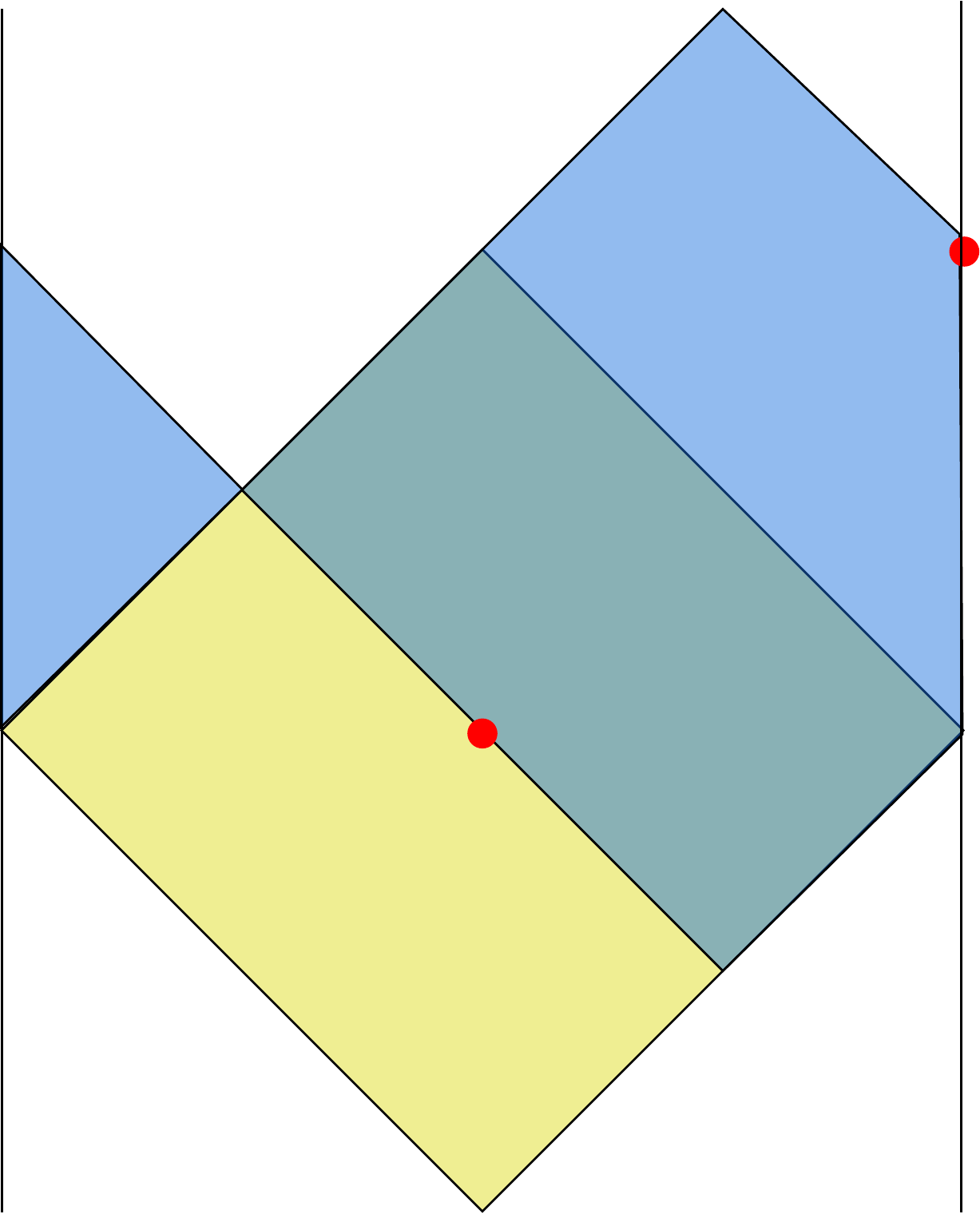}
\end{center}
\caption{\footnotesize 
Boundary version of figure \ref{fig:cylinder}, showing the action of the null inversion \eqref{fourshift} on Minkowski patches. The left and right sides of the diagram are identified to make the Lorentzian cylinder. The $(u',v')$ coordinates cover the Minkowski spacetime shown as a yellow diamond. The coordinates $(u,v)$ cover the shifted patch, shown as a blue diamond. These overlap in the region $v'>0$, $v < 0$. The shockwave hits the boundary at the red dots; it is created at the origin of the yellow diamond, which is on $I^-$ of the blue diamond.
\label{fig:shift}}
\end{figure}

So far we have been working in coordinates where the shockwave operator is inserted near the origin of Minkowski space.  Mapping the CFT to a Lorentzian cylinder, this Minkowski space covers just one patch --- the shaded yellow patch in figure \ref{fig:sphericalshock}. Let us now shift to the next patch using the null inversion
\cite{Hofman:2008ar}:
\be\label{fourshift}
v \rightarrow - \frac{z_0^2}{v} , \quad u \rightarrow u - \frac{\vec{x}\, {}^2}{v} , \quad x^i \rightarrow \frac{z_0 x^i}{v} \ ,
\ee
where $z_0$ is some length scale. This is the conformal transformation induced on the boundary by the AdS coordinate change \eqref{fourshift1}. The old and new Minkowski patches are shown on the cylinder in fig.~\ref{fig:shift}.

The origin in the old patch is a point on past null infinity in the new patch.  Therefore in these coordinates, we have a shockwave state $|\psi_\infty\rangle$ created by insertion of a local scalar operator near infinity:
\be\label{psiinf}
|\psi_\infty \rangle=\psi(x_0)|0\rangle\ , \qquad \langle\psi_\infty|=\langle 0| \psi(x_0^*)\ ,
\ee
where now
\be\label{points}
x_0=(u_0,v_0,\vec{x}_0)=(i\delta,\frac{i z_0^2}{\delta},\vec{0})\ , \qquad x_0^{*}=(-i\delta,-\frac{i z_0^2}{\delta},\vec{0})\ .
\ee
Note that $\langle \psi_\infty | \psi_\infty \rangle=(4z_0^2)^{-2\Delta_\psi}$. Again, the Hermitian conjugate is the standard one acting on states in Minkowski space.

This is the dictionary for the planar shockwave quoted in the introduction. Note that the stress tensor is real, as it must be, since it is an expectation value in the `in-in' sense.

To find the stress tensor, we can either apply \eqref{OP} directly to the new kinematics, or apply the null inversion to \eqref{st5}. The result agrees with the planar shockwave \eqref{gravityT} with energy $E = \Delta_\psi/(2\delta)$ in the limit $\delta \to 0$.

\subsubsection{Cylinder picture}
Although it may seem strange to insert a local operator at $v \sim i \infty$, it is natural when viewed on the global cylinder in figure \ref{fig:cylinder} or \ref{fig:shift}. To see this, set $\vec{x}=0$ and $z_0 = 1$. The map from the plane to the cylinder (in the sense of the Poincare patch embedding) is\footnote{Including the transverse directions,
\be
u,v = \frac{\sin \tau \mp \cos\phi\sin\theta}{\cos\tau - \cos\theta} \ , \quad
x_i  = \frac{\sin\theta \sin\phi \Omega_i}{\cos\tau - \cos\theta}
\ee
where $\Omega_{i=1\dots d-2}$ with $\Omega_i^2 = 1$ are coordinates on a unit $S^{d-3}$. Above we set $\phi = \pi$.
}
\be\label{uvcyl}
u = \frac{1}{\tan\frac{\theta-\tau}{2}} , \qquad v =- \frac{1}{\tan \frac{\theta + \tau}{2}} \ ,
\ee
where the cylinder coordinates are $-d\tau^2 + d\theta^2 + \sin^2\theta d\Omega_{d-2}^2$.  The null inversion, $v' = -1/v$, translates along the cylinder in the null direction:
\be
\theta' = \theta + \frac{\pi}{2} , \quad \tau' = \tau + \frac{\pi}{2}  \ .
\ee
To create the spherical shockwave, we insert $\psi$ at $u=v=i\delta \ll 1$, \ie 
\be
\theta = \pi \ , \qquad \tau = 2i\delta \ .
\ee
The planar shockwave \eqref{points} is inserted at
\be
\theta  = \frac{\pi}{2} , \qquad \tau = -\frac{\pi}{2} + 2 i \delta \ .
\ee
So we see that the insertion at $v = i \infty$ simply corresponds to a small Euclidean-time offset on the cylinder. 

Notice also that \eqref{uvcyl} is invariant under $(\tau,\theta) \to (\tau+\pi, \theta+\pi)$. In the $(u,v)$ patch, this takes points on $I^-$ to points on $I^+$. This will have interesting implications for positivity conditions in CFT with timelike separated points, discussed in section \ref{sss:tpositivity}.

\section{The Regge OPE}\label{s:reggeope}

\begin{figure}

\begin{center}
\usetikzlibrary{decorations.markings}    
\usetikzlibrary{decorations.markings}    
\begin{tikzpicture}[baseline=-3pt, scale=1.3]

\begin{scope}[very thick,shift={(4,0)}]
\coordinate (v1) at (-1.5,-1.5) {};
\coordinate(v2) at (1.5,1.5) {};
\coordinate (v3) at (1.5,-1.5) {};
\coordinate(v4) at (-1.5,1.5) {};

\draw[thin,-latex]  (v1) -- (v2)node[left]{$v$};
\draw[thin,-latex]  (v3) -- (v4)node[right]{$u$};
\draw[very thick,-latex,red]  (1.4,-1.4) -- (-0.2,0.2);
\draw[very thick,red] (0,0) --  (-1.35,1.35);
\draw  (0,3) -- (3,0);
\draw  (0,3) -- (-3,0);
\draw  (0,-3) -- (3,0);
\draw  (0,-3) -- (-3,0);
\coordinate(v5) at (0,0) {};
\draw(v5)node[right]{Shockwave};
\def \fac {.6};
\draw[
	scale=.5,samples=50,thick,blue,domain=2.5:4.6,variable=\y,
	postaction=decorate, 
	decoration={markings, 
		mark=at position 0 with {\draw[fill,black] circle (0.04) node[left,black]{$\psi$};},
		mark=at position 1 with {\arrow{>}},
		}]
	plot ({-\fac/(\y)-\fac*\y},{-\fac/(\y)+\fac*\y});
\draw[
	scale=.5,samples=50,thick,blue,domain=2.5:4.6,variable=\y,
	postaction=decorate, 
	decoration={markings, 
		mark=at position 0 with {\draw[fill,black] circle (0.04) node[right,black]{$\psi$};},
		mark=at position 1 with {\arrow{>}},
		}]
	plot ({\fac/(\y)+\fac*\y},{\fac/(\y)-\fac*\y});

\end{scope}

\end{tikzpicture}
\end{center}
\caption{\label{fig:regge} \small The Regge limit: the operator product $\psi(u,v)\psi(-u,-v)$ can be replaced by a shockwave propagating along $v=0$ when $\Delta_\psi \gg 1$ .}
\end{figure}
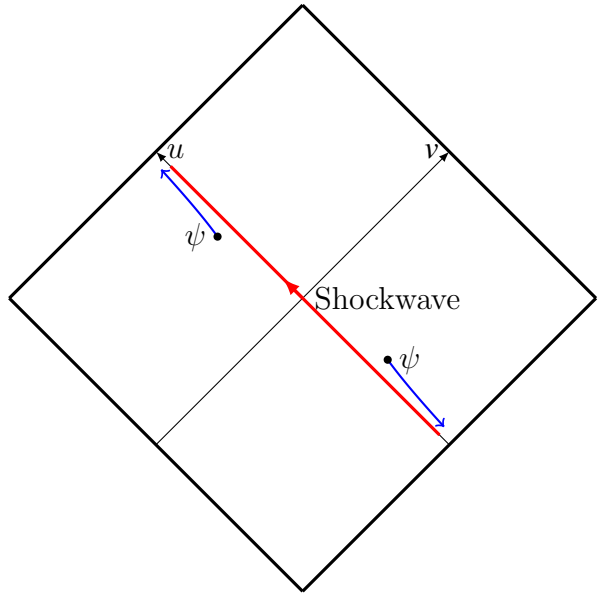

In the previous section, we found that  the spherical and planar shockwaves in AdS have exactly the same $\langle T_{\mu\nu}\rangle$ as the CFT states $|\psi_0\rangle$ and $|\psi_\infty\rangle$, respectively. This is a consequence of conformal symmetry that holds in any CFT; we have not yet used holographic duality. To claim that the states are dual requires $n$-point functions to match in this state as well. In a holographic CFT, finding the correct 1-point functions for single trace operators is enough, since these fully determine the bulk geometry, so higher-point functions are guaranteed to match. We will work it out explicitly to see how the shockwave naturally comes out of the OPE on the CFT side. 

Consider a scalar operator $\psi$ in the CFT, inserted symmetrically about the origin: $\psi(u,v)\psi(-u,-v)$, with $v < 0 < u$.  The Regge limit \cite{Brower:2006ea,Cornalba:2006xk,Cornalba:2006xm,Cornalba:2007zb,Cornalba:2007fs,Cornalba:2008qf,Costa:2012cb} of the OPE is defined by sending
\be\label{reggedef}
v \to 0 , \quad u \to \infty , \quad uv = \mbox{fixed} \ .
\ee
See figure \ref{fig:regge}.
This limit is usually discussed inside 4-point functions, but it also makes sense within the OPE, assuming that any other operator positions are held fixed at finite values as $v \to 0$. We will first derive the OPE on the gravity side, then compare to CFT. We assume that $\psi$ is a heavy probe operator, meaning $c_T \gg \Delta_\psi \gg 1$, where $c_T \sim 1/G_N$ is the coefficient of the stress tensor two-point function. The restriction to large $\Delta_\psi$ ensures that we do not need to worry about the $\psi$ 1-point function, so that the states discussed in section \ref{s:dictionary} are indeed dual to localized shockwaves.

(Caveat: The roles of $u$ and $v$ are swapped in this section compared to section \ref{s:dictionary}. Eventually we will need to discuss shockwaves going in both directions, so this is unavoidable.)

\subsection{The length operator}\label{length_op}

On the gravity side, the two-point function of a heavy probe can be computed in the WKB approximation. Assuming the bulk field dual to $\psi$ does not interact with any background fields that are turned on, it is given by the geodesic length connecting the two insertions,
\be
\langle\Psi| \psi(x_1) \psi(x_2)| \Psi\rangle =  e^{-\Delta_\psi L(x_1, x_2)} \ .
\ee
This holds in any state $|\Psi\rangle$ with a geometric dual, so we can attempt to write this as an operator relation:
\be\label{psil}
\psi(x_1)\psi(x_2) = e^{-\Delta_\psi L(x_1, x_2)} \ ,
\ee
where now $L$ is an operator built from the bulk metric. This is not a true operator equation, but holds at least in semiclassical states. The same relation has been exploited recently in the $d=2$ context \cite{Anous:2017tza,maxfieldtalk}, but here we restrict to $d \geq 3$.

By expanding \eqref{psil} perturbatively, we can turn this into a simple OPE. Choose $x_1 = -x_2 = (u,v,\vec{0})$, and write the bulk metric as
\be
g_{\mu\nu} = g_{\mu\nu}^{AdS} + h_{\mu\nu} \ .
\ee
In pure AdS, the geodesic that connects $x_1$ and $x_2$ is given by
\be
v'(u')=\frac{u' v}{u}\ , \qquad z'(u')=\sqrt{\frac{(u'^2-u^2)v}{u}}\ , \qquad \vec{x}'(u')=\vec{0}\ .
\ee
Now expanding the length operator to linear order in $h_{\mu \nu}$ yields
\be\label{ope1}
\frac{\psi(u,v)\psi(-u,-v)}{\langle \psi(u,v)\psi(-u,-v) \rangle}=  1 - \Delta_\psi \int_{-u}^u du' \frac{(u^2-u'^2)}{2u^3}\left(u^2 h_{uu}+ 2u v h_{uv}+ v^2 h_{vv}\right)+\O(h^2)\ ,
\ee
where $h_{\mu\nu}=h_{\mu\nu}(u',v'(u'),z'(u'),\vec{x}'(u'))$. In the Regge limit \eqref{reggedef} this becomes
\be\label{gravityope}
\frac{\psi(u,v)\psi(-u,-v)}{\langle \psi(u,v)\psi(-u,-v) \rangle}=  1 - \frac{\Delta_\psi u }{2}\int_{-\infty}^{\infty}du' h_{uu}(u',v'=0,z'=\sqrt{-uv}, \vec{0})\ .
\ee
This is the Regge OPE, written in gravity language, and is one of our main results in the simplest setting of heavy operator insertions. The integral is over a null geodesic parallel to the boundary,  \ie the source for a planar shockwave, so we refer to the $O(h)$ correction as a shockwave operator.  By exchanging $\Delta_\psi$ for $E$, the shockwave energy, it can also be written in the form \eqref{introres}.

We will make a few remarks on the bulk interpretation of \eqref{gravityope}, then turn to the CFT and examine its operator content.

\subsubsection{From Witten diagrams}\label{sss:witten}
The same formula \eqref{gravityope} can be derived from Witten diagrams. Consider the scalar-scalar-graviton vertex diagram in AdS:
\begin{equation}\label{vertexfig}
\usetikzlibrary{decorations.markings}    
\usetikzlibrary{decorations.markings}    
\begin{tikzpicture}[baseline=-3pt,scale=0.8]

\begin{scope}[very thick,shift={(4,0)}]

\draw (0,0) circle [radius=3];
\draw  (1.5,0) circle (1.5pt)-- (2.12,2.12);
\draw  (1.5,0) -- (2.12,-2.12);
\draw [red,domain=-1.5:1.5, samples=500] plot (\x, {0.1* cos(5*pi*\x r)});
\draw(2.5,2.12)node[above]{\large $\psi(x_1)$};
\draw(2.5,-2.12)node[below]{\large $\psi(x_2)$};
\draw(1.5,0)node[right]{ $(z,x)$};
\draw(-1.5,0)node[below]{ $(z',x')$};
\draw(0,0)node[above]{\large $h_{\alpha' \beta'}$};
\draw[red]  (-1.5,0) circle (1.5pt);
\end{scope}

\end{tikzpicture}
\end{equation}
In the Regge limit, with $\Delta_\psi \gg 1$, this diagram reduces to
\be\label{regge2}
\Pi_{\alpha' \beta'}(x_1,x_2;z',x')=-\frac{\Delta_\psi u}{2} \int_{-\infty}^{\infty} du'' G_{uu\alpha' \beta'}(u'',v=0, \vec{x}=0, z=\sqrt{-uv}; z',x')
\ee
where $G_{\alpha\beta\alpha'\beta'}$ is the graviton propagator in the bulk. This indicates that the full vertex is dominated by a single geodesic Witten diagram  \cite{Hijano:2015zsa}; the geodesic becomes null in the Regge limit, so it can be viewed as the source for a shockwave. The derivation of \eqref{regge2} is in appendix \ref{app:witten}. 

The vertex result \eqref{regge2} is equivalent to the OPE statement \eqref{gravityope}. To see this, simply insert both equations into a Witten diagram.

\subsubsection{Relation to imaginary shockwaves}\label{sss:relation}
Inserting the Regge OPE \eqref{gravityope} into a correlation function shows that we can replace the $\psi$ operators by a linearized shockwave:
\be\label{imcor}
\frac{\langle \psi(x_1) O_3(x_3) O_4(x_4) \cdots O_n(x_n)  \psi(x_2) \rangle}{\langle\psi(x_1)\psi(x_2)\rangle}  \approx \langle O_3(x_3)O_4(x_4)\cdots O_n(x_n) \rangle_{shock}
\ee
Here the $O$'s are primary operators, possibly with spin, obeying $\Delta_O \ll c_T$ so that we can work perturbatively. (Operator ordering is discussed below.) On the right-hand side is the $(n-2)$-point function in a shockwave background, with metric 
\be\label{imshock}
ds^2 = \frac{1}{z'^2}(-du' dv' + dz'^2 + d\vec{x}'^2) +  h_{vv}^{Shock} dv'^2
\ee
where $h_{vv}^{Shock}$ is the shockwave metric with $z_0 = \sqrt{-uv}$ and imaginary energy:
\be
E = \frac{i \Delta_\psi}{2v} \ .
\ee
(This shockwave is oriented in the opposite direction as \eqref{huu}, so $u \to v'$ in that formula, and we have set $L=1$.)

Note the crucial factor of $i$: If the $\psi$'s are inserted at real points $(u,v)$ in Minkowski spacetime, then the correlator \eqref{imcor} is computed in a purely imaginary shockwave.  The $i$ can also be seen by computing $\langle \psi(u,v) T_{\mu\nu}(0) \psi(-u,-v)\rangle$, which is purely imaginary for real $u,v$, indicating that the bulk metric perturbation is also imaginary.
This is related to the discussion in section \ref{s:dictionary}, where we found that the real shockwave corresponds to operators inserted at imaginary $(u,v)$. 

In terms of Witten diagrams, the same conclusions follow from doing the $u''$ integral in \eqref{regge2}, which gives the shockwave metric:
\be\label{vertexpi}
\Pi_{\alpha' \beta'}(x_1,x_2;z',x') = \tfrac{1}{2} \langle \psi(u,v)\psi(-u,-v)\rangle h_{vv}^{Shock} \delta_{\alpha'}^v \delta_{\beta'}^v \ .
\ee

\subsubsection{Real shockwaves}

The OPE, and the result \eqref{imcor}, also apply to $u \to i\infty$ as in the shockwave state $|\psi_\infty\rangle$ discussed in section \ref{s:dictionary}. This is the limit that must be taken in order to reproduce correlation functions in the real shockwave geometry, \ie the metric \eqref{huu} with $u \to v'$ and real $E = \Delta_\psi/(2\delta)$. This confirms, as expected, that higher-point functions in the shockwave state $|\psi_\infty\rangle$ indeed agree with gravity calculations on the real shockwave background.

\subsubsection{Operator ordering}\label{sss:ordering}

The operator ordering in \eqref{gravityope} is encoded in the choice of contour for the $u'$-integral. This is entirely analogous to the lightcone limit, discussed in detail in section 3 of \cite{Hartman:2016lgu}. Effectively, the $u'$ integral circles poles coming from operators inserted to one side of the $\psi\psi$ insertion in the correlators, and avoids poles from operators on the other side.  We will work out some explicit examples in section \ref{ss:timedelays}.

 \subsection{CFT interpretation}\label{s:cftreg}

 The Regge OPE \eqref{gravityope} is written in terms of the bulk metric. Next we want to interpret it in terms of CFT operators. At this point, we need to make the distinction between holographic and non-holographic CFTs. In any CFT, the stress tensor conformal block grows in the Regge limit, and we will see below that this growth is captured exactly by the shockwave operator \eqref{gravityope}. But this is not necessarily useful, because in general CFTs, the OPE cannot be used to calculate correlators in the Regge limit. Higher spin operators have increasingly large contributions, and there is little to be learned from the subleading stress tensor term.
 
 In a holographic CFT, on the other hand, with large-$N$ factorization of correlators and a large gap in the spectrum of higher spin operators, the Regge OPE is under control in the $1/N$ expansion. In this case, from the derivation, it is clear that \eqref{gravityope} includes anything that can be computed by a graviton exchange Witten diagram. In terms of CFT operators, this includes $T_{\mu\nu}$ itself, as well as double-trace operators $[\phi_1 \phi_2]$ built from all of the light operators in theory, including products of stress tensors $[TT]$. In a four-point function, it gives the dominant term in the Regge limit whenever the exchange diagram dominates. This breaks down deep into the Regge regime where the exchange of massive higher spin particles (e.g., string states) becomes important.

 \subsubsection{Single trace contribution}\label{sss:singleT}
 
Let's examine the single-trace contribution, which is universal to all CFTs. The Regge OPE \eqref{gravityope} is written in terms of the bulk metric. Using the HKLL prescription, we can rewrite it in terms of boundary CFT operators. Specializing to $d=4$, the HKLL formula for the metric perturbation is \cite{Kabat:2012hp}:
 \be\label{hkll}
 h_{\mu \nu}(t,y,\vec{x}, z)=\frac{8\pi G_N}{\pi^2 z^2}\int_{t'^2+y'^2+\vec{x}'^2<z^2}dt' dy' d^2 \vec{x}' T_{\mu\nu}(t+t',y+i y',\vec{x}+i \vec{x}') + \mbox{multitrace}\ .
 \ee
Plugging the single-trace term into \eqref{gravityope} and doing some of the integrals gives the stress tensor contribution to the Regge OPE:\footnote{Derivation:
The integral is
 \be\label{regge_hkll}
\frac{\psi(u,v)\psi(-u,-v)|_T}{\langle \psi(u,v)\psi(-u,-v)\rangle }=\frac{2 \lambda_T }{\pi^2 v}\int_{t'^2+y'^2+\vec{x}'^2<-uv}dt' dy' d^2 \vec{x}'  \int_{-\infty}^{\infty} du' T_{uu}\left(\frac{u'}{2}+t',-\frac{u'}{2}+i y',i \vec{x}'\right)\ .
 \ee
Now shift the contour $u' \to u' - t' + i y'$, and rewrite the integrals over $t',y'$ in the form $\frac{1}{2}\int_{|z|<-uv-t'^2}d^2 z T_{uu}(u=u', v=z, \vec{x} = i \vec{x}')$ where $z = t' + i y'$. Finally, do the complex $z$-integral by assuming the correlator is analytic within the disk of integration: $\int_{|z| \le R}  d^2z f(z) =2 \pi R^2 f(0)$. This assumption is valid in the four-point functions where we will apply \eqref{tblock}.}
\be\label{tblock}
\frac{\psi(u,v) \psi(-u,-v)|_T}{\langle \psi(u,v) \psi(-u,-v) \rangle } = \frac{2 \lambda_T u}{\pi}  \int_{-\infty}^{+\infty}  d\tilde{u} \int_{\vec{x}^2 \le - u v} d\vec{x}^2 \frac{ -uv  - \vec{x}^2}{u v} T_{uu} (\tilde{u} ,0 , \vec{i x} ) \ ,
\ee
where $\lambda_T=(2\pi G_N)\Delta_\psi=\frac{10 \Delta_\psi}{c_T \pi^2}$.

Although this was derived from gravity, it is completely general, for all CFTs and all values of $\Delta_\psi$. It is a statement about the piece of the stress tensor conformal block that grows in the Regge limit, so it can also be derived directly from CFT. This is most efficiently done using the OPE block formalism \cite{Czech:2016xec, deBoer:2016pqk}. The OPE block for a primary operator $O$ is the complete contribution of $O$ and its descendants to the $\psi\psi$ OPE.  Assuming temporarily that $x_1$ and $x_2$ are timelike separated, the OPE block for the stress tensor can be computed by integrating $T_{\mu\nu}$ over a codimension-1 spacelike surface  \cite{Czech:2016xec, deBoer:2016pqk}
\be\label{eq:formalope}
\psi(x_1) \psi(x_2) |_T =-\frac{2\lambda_T}{\pi^2}  \int_{B(x_1,x_2) } d\Sigma^{\mu} K^{\nu} T_{\mu \nu} \ .
\ee
Here $B(x_1,x_2)$ denotes a Cauchy surface within the causal diamond $future(x_1) \cap past(x_2)$, which we take to be the ball equidistant from $x_1$ and $x_2$.  The conformal Killing vector $K^\mu = \frac{-2\pi}{(x_2-x_1)^2} [(x_2 -x)^2 (x_1 -x)^{\mu} - (x_1-x)^2 (x_2 -x)^\mu]$  is the generator of time translations within the diamond. 

For spacelike separated $x_1$, $x_2$, the OPE block is obtained from \eqref{eq:formalope} by analytic continuation.  (Note that this immediately gives the Euclidean OPE block without `shadow' contributions.) In our kinematics, with $v<0<u$, the full OPE block is
\begin{align}\label{eq:tope}
\frac{\psi(u,v) \psi(-u,-v)|_T}{\langle \psi(u,v) \psi(-u,-v) \rangle } &= \frac{2 \lambda_T}{\pi}  \int_{-u}^u  d\tilde{u} \int_{\vec{x}^2 \le - v u \left(1- \frac{\tilde{u}^2}{u^2}\right)} d^2\vec{x} \frac{ -uv \left(1-  \frac{\tilde{u}^2}{u^2} \right) - \vec{x}^2}{ v} \\
& \times \left[ T_{uu} \left(\tilde{u}, - \frac{ v \tilde{u}}{u},i \vec{  x}\right) - 2\frac{v}{u} T_{vu} \left(\tilde{u}, - \frac{ v \tilde{u}}{u},i \vec{ x}\right) + \frac{v^2}{u^2} T_{vv} \left(\tilde{u}, - \frac{ v \tilde{u}}{u},i\vec{  x}\right)\right]\ .\notag
\end{align}
In the Regge limit \eqref{reggedef}, the ball integral becomes an integral over a slab on the null sheet $v=0$, and reproduces exactly the gravity result \eqref{tblock}.

In terms of 4-point conformal blocks, \eqref{tblock} captures precisely the growing part of the block in the Regge limit. For example, in $d=4$, the stress tensor conformal block has a growing part $\sim \frac{i\bz}{z(z-\bz)}$, where $z,\bz$ are the conformal cross ratios.  Inserting the operator \eqref{tblock} into the four-point function reproduces exactly this term, and no more. This requires some care with the contour of integration and is worked out explicitly in appendix \ref{app:doblock}. 

In the lightcone limit, which is $v \to 0$ with $u$ held fixed, the domain of integration for transverse part of the OPE block is small, and the ball integral becomes a line integral. Thus, we can set $\vec{x} =0$ in the argument of the stress tensor and  \eqref{eq:tope} reduces to the lightcone OPE derived in \cite{Hartman:2016lgu},
\be
\frac{\psi(u,v) \psi(-u,-v)|^{\text{lightcone}}_T}{\langle \psi(u,v) \psi(-u,-v) \rangle }  = \lambda_T  v u^2 \int_{-u}^u d\tilde{u} \left(1- \frac{\tilde{u}^2}{u^2}\right)^2 T_{uu} \left(\tilde{u},0,\vec{0}\right)  \ .
\ee
Again this agrees with the length operator \eqref{ope1} upon using HKLL to replace $h_{uu} \to T_{uu}$.

The analysis is similar in general dimensions, leading to the Regge OPE
\begin{align}
&\frac{\psi(u,v) \psi(-u,-v)|_T}{\langle \psi(u,v) \psi(-u,-v) \rangle }  =\notag \\
&\qquad \frac{\Delta_\psi 2^d \pi^{\half - d}\Gamma(\frac{d+2}{2})\Gamma(\frac{d+3}{2})}{c_T (d-1)}
u  \int_{-\infty}^{+\infty}  d\tilde{u} \int_{\vec{x}^2 \le -u v} d^{d-2}\vec{x} \frac{ -uv  - \vec{x}^2}{ uv} T_{uu} \left(\tilde{u} ,0 , i \vec{x} \right)\ .
\end{align}
As noted above, this is just the single-trace contribution from the operator $T_{\mu\nu}$ and its derivatives. We will see below that this term has a direct interpretation in terms of a time delay in the shockwave geometry.

\subsubsection{Regge OPE for other operators}
Although we will not use it anywhere in this paper, the same analysis can be applied to other operators. For the production of an operator $X$ with spin $\ell$ and dimension $\Delta$, the full OPE block can be written as an integral over the causal diamond between $x_1$ and $x_2$ \cite{Czech:2016xec, deBoer:2016pqk}, and analytically continued to spacelike separation. In appendix \ref{app:generalregge} we take the Regge limit and find
\begin{align}
\begin{split}\label{genreg}
& \frac{\psi(u,v) \psi(-u,-v)|_X}{\langle \psi(u,v) \psi(-u,-v) \rangle } =  \pi^{\frac{1 -d}{2}} 2^{\Delta}  \frac{\Gamma(\frac{\Delta+\ell+1}{2})}{\Gamma(\frac{\Delta+\ell}{2})} \frac{\Gamma(\Delta -d/2+1)}{\Gamma(\Delta-d+2)} \frac{C_{\psi \psi X}}{C_X} \\
&\times \frac{(-u v)^{\frac{d-\ell-\Delta}{2}}}{u^{1-\ell}} \int_{-\infty}^{+\infty} d\tilde{u} \int_{\vec{x}^2 \le -u v}d^{d-2}\vec{x} ( -u v -\vec{x}^2)^{\Delta-d+1} X_{uu\cdots u}(\tilde{u},0,i\vec{ x})
\end{split}
\end{align}
This reproduces the growing part of the 4-point conformal block.

\section{Causality constraints revisited}\label{s:causality}

\subsection{The Engelhardt-Fischetti criterion}

Causality is subtle in a theory with gravity, because local light cones can always be `opened' by a diffeomorphism. At the least, causality should be respected at infinity, in the sense that a probe sent from infinity that passes through some geometry should return to infinity no faster than it would in vacuum \cite{Penrose:1993ud,Woolgar:1993zp,Gao:2000ga}. That is, all time delays should be non-negative. In a geometry that is asymptotically AdS, the statement is that lightcones on the conformal boundary must be respected by the bulk. In AdS/CFT, the same criterion is imposed by causality of the dual CFT \cite{Kleban:2001nh}.

However, it remains an open question exactly what feature of the bulk theory ensures boundary causality. Gao and Wald showed that in Einstein gravity, the averaged null energy condition for bulk matter implies boundary causality \cite{Gao:2000ga}. More generally, it is respected in any theory of gravity obeying the averaged null curvature condition, $\int du R_{uu}  \geq 0$. But is this condition also necessary?  Perhaps not --- there are geometries which violate the curvature condition, but preserve boundary causality \cite{Engelhardt:2016aoo}. It is not known whether such solutions can be supported by physical matter.  Geometrically, the necessary and sufficient condition for boundary causality, at the perturbative level, was derived by Engelhardt and Fischetti \cite{Engelhardt:2016aoo}. It is simply the condition 
\be\label{efc}
\int_\gamma d\lambda h_{\lambda\lambda} \geq 0 \ ,
\ee
where the integral is over a complete null geodesic $\gamma$ in AdS. The integral computes a time delay, so the same condition has been imposed in specific states in prior work (for example \cite{Brigante:2007nu,Brigante:2008gz,Hofman:2008ar,Hofman:2009ug, Camanho:2014apa,Papallo:2015rna}). Any such null geodesic can be viewed as a constant-$z$ path in some Poincare patch.

The quantity in \eqref{efc} is precisely the operator that appears in the Regge OPE \eqref{gravityope}.   And, the chaos bound \cite{Maldacena:2015waa} implies that this contribution to the correlator, including the overall minus sign, must be negative. Therefore the chaos bound, together with the Regge OPE, constitutes a derivation of \eqref{efc}, now viewed as an expectation value of this operator in the quantum theory. (See \cite{Afkhami-Jeddi:2016ntf} for more discussion on the connection between chaos and the Regge limit of the 4-point function.) This is the Regge analogue of the averaged null energy condition (ANEC) as derived from causality in \cite{Hartman:2016lgu}.

Unlike the ANEC, the conclusion here comes with the caveat that we have only shown \eqref{efc} is positive in a certain class of states. Exactly which states depends on the assumptions, and how we interpret the formula:
\renewcommand{\theenumi}{$(\roman{enumi})$}
\begin{enumerate}
\item If $h_{uu}$ is interpreted as the full metric perturbation operator, with contributions from multitrace operators as well as the stress tensor, then the conclusion is that the chaos bound implies \eqref{efc} in any holographic CFT dual gravity, in any state perturbatively close to the vacuum.  By `gravity', we mean that the bulk theory has a large gap without any single trace higher spin states, but may have large higher curvature corrections. The state does not need to be geometric; for example, it could be a superposition of classical geometries, with $h_{uu}$ evaluated as an expectation value.  In general, the multitrace contributions are important, so in CFT language, this inequality would be quite complicated, and cannot be written in terms of just the operator $T_{\mu\nu}$. 
\item In certain states, single-trace exchange dominates, and we can replace $h_{uu} \to h_{uu}|_T$, the single-trace, stress-tensor contribution to HKLL. These are states $|\phi\rangle$ in which $\langle \phi|\psi\psi|\phi\rangle$ is dominated by the stress tensor conformal block. In a large-$N$ CFT with a large gap in the higher spin spectrum, this includes states created by a heavy (but not backreacting) operator insertion $\phi(x)$, with $c_T \gg \Delta_\phi \gg 1$, as well as superpositions of such states, and states with multiple heavy operator insertions.  We will show below that it also includes certain states produced by smeared light operators.
\end{enumerate}
In gravity language, we have shown that the geodesic time delay is positive in any state near the vacuum.

So far, this argument is not enough to reproduce the ``$a=c$" constraints derived from graviton causality in \cite{Camanho:2014apa}. We will derive these below, by projecting out double trace contributions to correlators of light operators.  In the rest of this section, we focus on the case of 4 heavy scalars, and describe how the Regge OPE is directly related to a bulk time delay.

\subsection{Time delays for probe particles}\label{ss:timedelays}

To make this more concrete, we will now compare two different calculations of a 4-point correlator: The CFT calculation using the Regge OPE, and the bulk calculation using geodesic lengths in the shockwave background. The point is that $\Delta v$, the time delay as a scalar particle crosses a shock, is computed precisely by the Regge OPE  for the stress tensor alone, \eqref{tblock}. This is almost clear from the gravity formula \eqref{gravityope} but there are some subtleties in the prescription for how to apply this OPE, especially as the probe geodesic moves past the shockwave source into the bulk.

\subsubsection{Setup}\label{sss:tdsetup}

We consider the scalar four-point function
\be\label{defgg}
G  \equiv \frac{\langle \psi(x_1)\phi(x_3)\phi(x_4)\psi(x_2)\rangle}{\langle \psi(x_1)\psi(x_2)\rangle\langle \phi(x_3)\phi(x_4)\rangle} \ .
\ee
There are two cases where the stress tensor term \eqref{tblock} is the dominant contribution, without any contamination from double trace or other operators. One is the lightcone limit, in any CFT, where it dominates because $T_{\mu\nu}$ is the operator of lowest twist. The other is the Regge limit in a large-$N$, holographic CFT, dual to  gravity (possibly with higher curvature corrections but without higher spin fields), with all four operators heavy: $c_T \gg \Delta_{\psi}, \Delta_\phi \gg 1$. 

We'll start with the Regge limit. In this case the OPE is controlled by $1/N$, and double trace contributions from $[\psi\psi]$ and $[\phi\phi]$ are suppressed by their large dimensions. In order to utilize the shockwave dictionary \eqref{introdict}, with a real shockwave geometry, we choose the imaginary kinematics
\be\label{geokina}
x_1 = -x_2 \equiv (u_\psi, v_\psi, \vec{x}_\psi ) \equiv  (i\delta,\  \frac{i z_0^2}{\delta},\  \vec{0}) \ .
\ee
The points $x_1$, $x_2$ are conjugate, so $G$ is an expectation value:\footnote{Note that these are not the kinematics used in the derivation of the chaos bound. The sign constraints discussed in section \ref{s:causality} are derived using kinematics where $x_1$ and $x_2$ are related by a reflection across the Rindler horizon, rather than a Hermitian conjugation \cite{Maldacena:2015waa} (see also \cite{Afkhami-Jeddi:2016ntf}). However the chaos bound does imply --- indirectly --- that $\Delta v$ in the current calculation must be positive, because it is computed by the same operator, which must have positive expectation value in any state. In order to make a more direct connection between the chaos argument and the bulk time delay would require an imaginary metric in the bulk.}
\be\label{gexp}
G = \frac{\langle \psi_\infty| \phi(x_3)\phi(x_4) |\psi_\infty\rangle}{\langle \psi_\infty|\psi_\infty\rangle\langle \phi(x_3)\phi(x_4) \rangle} \ .
\ee
The shockwave lies on the Rindler horizon, at $u = 0$. We also insert $\phi$ symmetrically across the Rindler wedge:
\be\label{phikina}
x_3 = (u_\phi, v_\phi, \vec{x}_\phi ) \ , \quad
x_4 = (-u_\phi, -v_\phi, \vec{x}_\phi ) \ , \quad \mbox{with} \quad v_\phi < 0 < u_\phi\ .
\ee
The limit $\delta \to 0$ is taken with $u_\phi, v_\phi, \vec{x}_\phi, z_0$ fixed and $O(1)$. 

\begin{figure}[t]
\begin{center}
\includegraphics[width=3in]{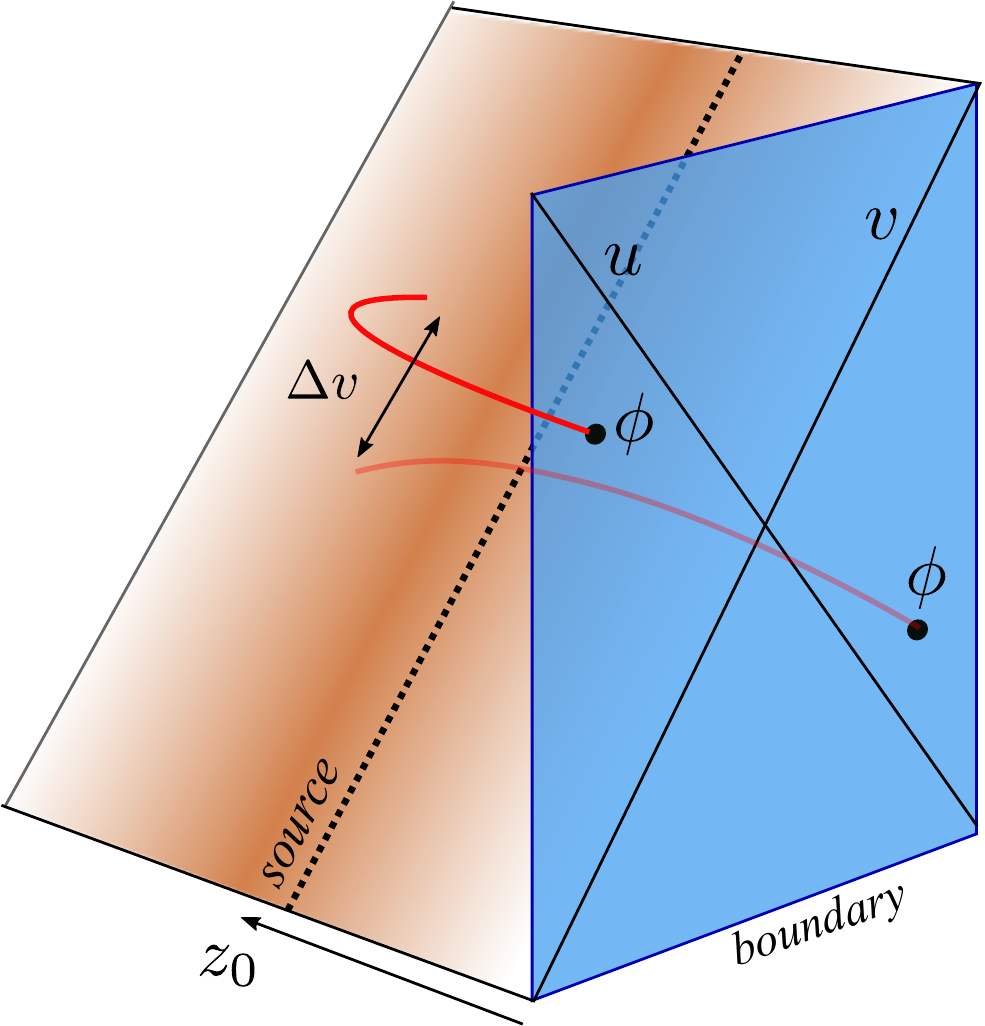}
\end{center}
\caption{\footnotesize 
Setup for the four-point function. The metric is a real shockwave, created by $\psi$ inserted at imaginary kinematics. $\phi$ is a probe, and $\langle \psi| \phi \phi| \psi\rangle$ is computed from the geodesic length. The geodesic jumps by $\Delta v$ across the shockwave. The transverse directions $\vec{x}$ also play a role but are suppressed in the figure.
\label{fig:geodesicsetup}}
\end{figure}

This setup is illustrated in figure \ref{fig:geodesicsetup}.

\subsubsection{Geodesics in the shockwave}\label{sss:tdgeo}
On the gravity side, $|\psi_\infty\rangle$ is the planar shockwave with metric \eqref{swmetric} and $E = \Delta_\psi/(2\delta)$, so $G$ is the two-point function $\langle \phi \phi\rangle_{shock}$ in this background. Since $\phi$ is heavy, it is computed by a geodesic length:
\be\label{geog}
G = \frac{\langle \phi(x_3)\phi(x_4)\rangle_{shock}}{\langle \phi(x_3)\phi(x_4)\rangle_{vacuum}} = \exp\left[-\Delta_\phi (L_{shock}(x_3, x_4) - L_{vac}(x_3,x_4))\right] \ .
\ee
The metric is empty AdS away from the shock. When the geodesic crosses the shockwave at $u=0$, heading in the positive-$u$ direction, it jumps by $v \to v + \Delta v$, with
\be\label{defdv}
\Delta v(z_c, \vec{x}_c) = \left. z^2 h \right|_{z = z_c, \vec{x} = \vec{x}_c}
\ee
where $(z_c, \vec{x}_c)$ are coordinates at the crossing point, and $h$ is given by $h_{uu}^{Shock}$ in \eqref{swmetric} with the delta function stripped off, \ie $h = \int_{0^-}^{0^+}du h_{uu}$. In principle, finding the full geodesic is just a matter of patching together empty-AdS geodesics with the correct endpoints. This is somewhat complicated because crossing the shockwave also imparts transverse $\vec{x}$-momentum to the probe, but we only need the first perturbative $O(E)$ term, which is straightforward. The calculation is exactly the same as that of section (\ref{length_op}), yielding 
\be\label{ggnice}
G \approx 1-\frac{\Delta_\phi u_\phi}{2} \int_{-u_\phi}^{u_\phi} du' \left(1-\frac{u'^2}{u_\phi^2} \right) h_{uu}^{Shock}=1 - \frac{\Delta_\phi u_\phi}{2} \frac{\Delta v(z_c, \vec{x}_c)}{z_c^2}
\ee
with the crossing point at
\be
z_c = \sqrt{-u_\phi v_\phi} , \quad \vec{x}_c = \vec{x}_\phi \ .
\ee
For example, in $d=4$ with $\vec{x}_\phi = 0$, the time delay \eqref{defdv} in the metric \eqref{swmetric} is
\be
\Delta v = 8 E G_N \times \begin{cases}
\frac{ z_c^4}{z_0^2 - z_c^2} & z_c < z_0 \\
\frac{ z_0^4}{z_c^2 - z_0^2} & z_c > z_0 
\end{cases} \ .
\ee
Using $E = \Delta_\psi/(2\delta)$,  $c_T = \frac{5}{\pi^3 G_N}$, and \eqref{geokina}, this becomes
\be\label{g4final}
G - 1 \approx - \frac{10 \Delta_\phi \Delta_\psi}{\pi^3 c_T|v_\phi u_\psi| }  \times  \begin{cases}
\frac{|u_\phi v_\phi|^2}{|u_\psi v_\psi| - |u_\phi v_\phi|} 
& |u_\psi v_\psi| > |u_\phi v_\phi |\\
\frac{|u_\psi v_\psi|^2}{|u_\phi v_\phi| - |u_\psi v_\psi|} & |u_\psi v_\psi| < |u_\phi v_\phi|
\end{cases}
\ee
Here we can see the symmetry between the roles of the shockwave source $\psi$ and probe field $\phi$, as evident from \eqref{defgg}. It does not matter which pair is taken into the shockwave limit, since the two options are conformally equivalent. The non-analyticity at $u_\psi v_\psi = u_\phi v_\phi$ in \eqref{g4final} is where the probe geodesic hits the shockwave source, and is smoothed out by any nonzero transverse separation $\vec{x}_\phi - \vec{x}_\psi$. 

The general answer, at non-zero $\vec{x}_\phi$, can be written using the conformal cross ratios $z,\bz$. In the Regge limit,\footnote{Both cross ratios are of the form $z, \bz \sim -i\times(positive)$, and we choose (arbitrarily) the solution of the quadratic equation such that $\bz/z \to 0$ in the lightcone limit $v_\phi \to 0^-$. This implies $\bz /z$ is real and positive for any choice of external points within the regime considered here.}
\bea
z \bz  &=&\frac{x_{12}^2 x_{34}^2}{x_{13}^2 x_{24}^2}  \approx \frac{16 u_\psi v_\phi}{u_\phi v_\psi}\\
z + \bz &\approx& 1 - \frac{x_{14}^2 x_{23}^2}{x_{13}^2 x_{24}^2} = 4\frac{\vec{x}_\phi^2 + z_0^2 + z_c^2}{u_\phi v_\psi}
\eea
The variable $\rho$ that appears in $h_{uu}^{shock}$, evaluated at the crossing point, is
\be
\rho_c =\sqrt{\frac{(z_c-z_0)^2+\vec{x}_\phi ^2}{(z_c+z_0)^2+\vec{x}_\phi ^2}} = \frac{1 - \sqrt{\bz/z}}{1 + \sqrt{\bz/z}} \ .
\ee
Therefore, in $d=4$, 
\be\label{gzz}
G  \approx1+ \frac{40 \Delta_\phi \Delta_\psi}{\pi^3 c_T} \frac{i \bz}{z(z-\bz)}
\ee
Note that this single expression (which we will see is equal to a Regge conformal block) accounts for both of the piecewise answers in \eqref{g4final}. The jump comes from the fact that in the limit $\vec{x}_\phi \to 0$, the cross ratios (defined to be smooth at finite $\vec{x}_\phi$) become non-analytic in the  coordinates:  $z/\bz = \max(z_0^2/z_c^2, z_c^2/z_0^2)$. 

We can also make a direct connection between the correlator and the bulk equations of motion. Since the correction to the correlator was just the bulk metric $h_{uu}^{Shock}$ with a delta function stripped off, it obeys the Einstein equation, integrated over $u$:\footnote{The differential operator on the left is the conformal Casimir operator in the Regge limit, $z^2\p^2 + \bz^2\p^2 + \frac{2z\bz}{z-\bz}(\p - \bar{\p})$. Interestingly, acting on the block it produces a delta function source in Lorentzian signature.   It may be fruitful to understand the role of these delta functions in the conformal bootstrap.}
\be
\left(\partial_{z_c}^2-3\frac{\partial_{z_c}}{z_c}  +\partial_{\vec{x}_\phi}^2\right)
\left(\frac{\pi^2 c_T |v_\phi u_\psi|}{20 \Delta_\phi \Delta_\psi} \right)(G-1)
=z_0^3\delta(z_c-z_0)\delta(\vec{x}_\phi - \vec{x}_\psi)\ .
\ee

Although we have assumed Einstein gravity at intermediate steps, the final answer \eqref{gzz} written in terms of $c_T$ is also valid in the presence of higher derivative corrections (both $E$ and $G_N$ get rescaled, but $E G_N$ is unchanged).

\subsubsection{CFT calculation using the Regge OPE}\label{sss:timedelaycft}
Let's reproduce this from CFT.  One way to proceed is to use the Dolan-Osborn expression for the stress tensor conformal block; upon analytic continuation to the Regge regime, this gives directly \eqref{gzz}, as shown in appendix \ref{app:doblock}.  Instead we will use the Regge OPE \eqref{gravityope} to illustrate how to apply it. Since $\psi$ is the operator that is inserted in the shockwave limit, the direct approach would be to use \eqref{gravityope} in the form 
\be\label{psiope}
\psi\psi \sim 1 - \frac{\Delta_\psi v_\psi}{2} \int_{-\infty}^{\infty} dv h_{vv}|_T , 
\ee
and then to calculate $\langle \phi \int dv h_{vv} \phi\rangle$. The notation $|_T$ indicates that only the single-trace stress tensor part of $h_{\mu\nu}$ is included (via HKLL as discussed in section \ref{sss:singleT}). This will be useful below, but first, for comparison to the gravity result \eqref{ggnice} it is more informative to contract the probes $\phi\phi$ rather than the sources $\psi\psi$. That is, we use
\be\label{phireg}
\phi \phi \sim 1 - \frac{\Delta_\phi u_\phi}{2} \int du h_{uu}|_T \ .
\ee
Despite that fact that we have not taken $v_\phi, 1/u_\phi \to 0$, this OPE can be used because in the conformal cross-ratio it does not matter which operator is the source, and which is the probe. (In other words, we could boost so that $\phi\phi$ are Regge-separated with $\psi\psi$ near the origin, apply the Regge OPE to $\phi\phi$, then boost back.)

Formally, plugging \eqref{phireg} into the state $|\psi_\infty\rangle$ gives a correlator manifestly equal to the gravity result \eqref{ggnice}, since $\Delta v = \int du z^2 h_{uu}$. But to confirm this by direct evaluation in the CFT, we need to account for some subtleties involving the choice of $u$-contour. Converting \eqref{phireg} to a CFT expression using \eqref{tblock}, then evaluating in the shockwave state (with $\vec{x}_\phi = 0$, $d=4$) we find: 
\bea\label{dgcft}
G-1 &=& - \frac{20 u_\phi \Delta_\phi}{\pi^3 c_T} \int du  \int_{\vec{x}^2 < |u_\phi v_\phi|} d^2\vec{x} \frac{|u_\phi v_\phi| -\vec{x}^2}{|u_\phi v_\phi|}\langle \psi_\infty| T_{uu}(u, v=0, i \vec{x})|\psi_\infty\rangle  / \langle \psi_\infty|\psi_\infty\rangle \notag\\
&=& \frac{320\delta^4 \Delta_\phi \Delta_\psi z_0^6 }{3\pi^4 c_T |v_\phi|}\int du \int_0^{z_c} dr  \frac{r(r^2 - z_0^2)^2 (r^2 - z_c^2)}{(u^2 z_0^4 + (r^2-z_0^2)^2\delta^2)^3}
\eea
where on the second line we plugged in the conformal three-point function \eqref{OP}, and $z_0 = |v_\psi u_\psi|^{1/2}, z_c = |v_\phi u_\phi|^{1/2}$.
For $z_c < z_0$, the integrals are straightforward, with the $u$ integral done first. The correct prescription for the $u$ integral can be derived by boosting the $\phi$'s, integrating $\int_{-\infty}^\infty du$, then boosting back; the upshot is that the integral just picks up a residue.  After doing the $r$-integral, the final result is equal to the first line of \eqref{g4final}. Therefore, we have exactly reproduced the case where the probe geodesic crosses the shockwave closer to the boundary than the source particle.

In the other case, where the geodesic probe crosses the shockwave past the source, \ie $z_c > z_0$, the integral in \eqref{dgcft} diverges. The easiest way around this obstacle is to swap the role of source and probe, \ie to use \eqref{psiope} instead. Clearly this reproduces the gravity result on the second line of \eqref{g4final}, since it differs from the first line just by swapping the two operators.

Therefore, we have reproduced the full gravity answer using the Regge OPE. 

\subsubsection{Lightcone limit}

Everything in sections \ref{sss:tdsetup} - \ref{sss:tdgeo} can also be applied in the lightcone limit, where $u \to 0$ with $v$ held fixed. In this case, there is perfect agreement in all CFTs, regardless of whether they have a holographic dual, since the entire calculation happens near the boundary of AdS. On the CFT side, the perturbative analysis is valid because instead of $1/N$, the corrections to the correlator are suppressed by positive powers of $u$. On the bulk side, the perturbative geodesic calculation is valid because now $z_c \ll z_0$, and in that limit, the discontinuity in the geodesic where it hits the shockwave is also suppressed by $u$.  We omit the details but the result in the end is just the lightcone limit of \eqref{g4final}, in both boundary and CFT. This result demonstrates how log terms in the lightcone conformal block are related to time delays \cite{causality1}, and provides a precise equivalence between two different derivations of the averaged null energy condition: The holographic derivation \cite{Kelly:2014mra} using geodesic lengths, and the causality derivation use the lightcone OPE  \cite{Hartman:2016lgu}. 

\section{Shockwaves from light operators}\label{s:light}

So far, all of our discussion of shockwaves has required the insertion of two heavy operators $\psi$, with $\Delta_\psi \gg 1$. From a bulk point of view, heavy operators are massive probe particles that travel on geodesics. Now we turn to another way to make shockwaves, using wavepackets of light operators, designed to travel on geodesics in the bulk. We mostly restrict to $d=4$ for simplicity, but the results should generalized straightforwardly to any $d>2$.

\subsection{Smeared operators}\label{smearing}
The smearing procedure that we will use to produce shockwaves is identical, up to a conformal transformation, to that used in \cite{Afkhami-Jeddi:2016ntf}. There, it was motivated by a sequence of complexified rotations, shifts, and limits from the kinematics of Hofman and Maldacena \cite{Hofman:2008ar}. Here we work in a different conformal frame where the smearing is simpler to understand using a null inversion.

The starting point is an operator inserted at some fixed, real Lorentzian time $t = t_0 > 0$, smeared over complex positions with a Gaussian profile:
\be
W'(t_0) = \int dy d^{d-2}\vec{w} e^{-(y^2 + \vec{w}^2)/D^2} \phi(t_0,  iy, i\vec{w}) \ .
\ee
We implicitly cutoff the Gaussian outside some window, so the smearing has finite support; details off the cutoff won't matter, and the dominant contributions to the integrals will always come from a region where the Gaussian factor can be ignored. One can also think of this operator as a momentum space insertion, Wick rotated in all directions:
\be
\int d\tau d^{d-1} x\,  e^{-(\tau^2 + x^2)/D^2} e^{-i \omega \tau}\phi(t=i\tau, ix) \ .
\ee
This is not quite the same operator, but is equal inside the correlation functions we consider if evaluated at the saddlepoint $t = t_0 \sim  \omega D^2$.

Now perform the null inversion \eqref{fourshift}. The smeared operator (with the arguments of $\phi$ written in null coordinates $u,v$) becomes
\be\label{wdef}
W(t_0) = \int dy d\vec{w} e^{-(y^2 + \vec{w}^2)/D^2}
\left(t_0 + i y \over z_0\right)^{-\Delta_\phi} \phi(u, v, \vec{x} )
\ee
where under the integral, the coordinates are parameterized as 
\be\label{intco}
u = t_0 - i y  + \frac{\vec{w}^2}{t_0+iy}  \ , \qquad
v = -\frac{z_0^2}{t_0+iy} \ , \qquad
\vec{x} = i \frac{z_0 \vec{w}}{t_0+iy} \ .
\ee
The center of the smeared operator,  $y = \vec{w} = 0$, is
\be
u_{center} = t_0 , \qquad v_{center} = - \frac{z_0^2}{t_0}  \ .
\ee
Therefore the Regge limit is $t_0 \to 0$. The limit is taken with $D/t_0$ fixed but large, so that all of the coordinates $(t_0, y, \vec{w})$ effectively scale together towards zero.

We will insert a second wavepacket which is obtained from $W$ by sending $t_0 \to -t_0$.  Importantly, after the null inversion, this is equivalent to reflecting across the Rindler horizon, up to a fixed phase. Rindler reflection sends
\be
(u,v,\vec{x}) \to (\bar{u}, \bar{v}, \overline{\vec{x}}) \equiv (-u^*, -v^*, \vec{x}^*) \ .
\ee
Thus we define
\be
\overline{W}(t_0) \equiv (-1)^{\Delta_\phi} W(-t_0) = \int dy d\vec{w} e^{-( y^2 + \vec{w}^2)/D^2}\left(t_0 - i y \over z_0\right)^{-\Delta_\phi}\phi(-u^*, -v^*, \vec{x}^*) \ ,
\ee
where $u,v,\vec{x}$ are given by \eqref{intco}.The phase comes from the conformal factor in the null inversion. 

Rindler reflection is important for positivity/causality constraints, because Rindler-symmetric correlators have properties akin to reflection positivity in Euclidean signature; see \cite{Casini:2010bf,Maldacena:2015waa,Hartman:2016lgu} for a detailed discussion.

\subsubsection{Positivity at timelike separation}\label{sss:tpositivity}
As an aside, note that under the null inversion, the fact that Rindler-symmetric correlators are positive leads to a positivity condition for timelike separated insertions. The simplest example of this positivity condition is the vacuum two-point function, $\langle O(x_1)O(x_2)\rangle = (x_{12})^{-2\Delta}$. With real $t$ and any $x$, this obviously obeys the positivity condition
\be\label{genpp}
(-1)^{\Delta} \langle O(t, x) O(-t, x)\rangle = |2t|^{2\Delta} > 0 \ .
\ee
Rindler positivity, after a null inversion, implies that a similar relation holds for $n$-point functions, with a specific phase dictated by the conformal weights.  This is somewhat surprising --- it is a positivity condition in unitary CFTs that applies to timelike separate insertions. In what follows, we will work in the coordinates where operators are inserted symmetrically across the Rindler horizon, so chaos/causality bounds are a consequence of Rindler reflection positivity. Alternatively, one can work directly in the $(t,y, \vec{w})$ patch, with the shockwave inserted near the origin; in this approach the chaos/causality bounds follow from a timelike positivity condition like \eqref{genpp}.

\subsubsection{Cross ratios}

\begin{figure}[t]
\centering
\includegraphics{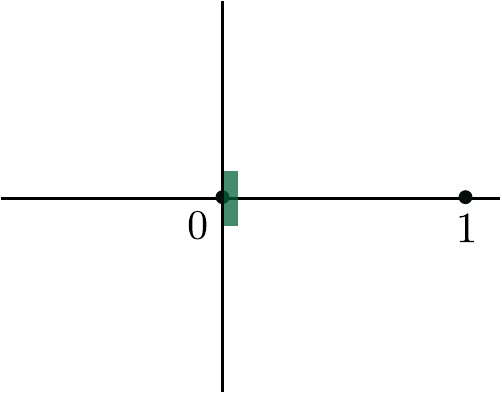}
\caption{\footnotesize
The smearing region on the $z$ and $\bz$ planes is the shaded rectangle near the origin. The Regge limit is $z,\bz \to 0$.
\label{fig:zregion}
}
\end{figure}

Consider these wavepackets inserted in a 4-point function,
\be
G  = \langle \overline{W}(t_0) W(t_0) \phi'(x_3) \phi'(x_4)\rangle \ .
\ee
This is a double integral of $\langle \phi(x_1) \phi(x_2) \phi'(x_3) \phi'(x_4) \rangle$, where
\bea
x_1 &=& (u_1,v_1, \vec{x}_1) \\
x_2 &=& (-u_2^*, -v_2^*, \vec{x}_2^*)
\eea
with $u_i,v_i, \vec{x}_i$ defined as in \eqref{intco}. Let's choose the other operators to also be symmetric across the Rindler horizon,
\be
x_3  = -x_4 = (u=-1, v=1, \vec{0}) \ .
\ee
The conformal cross-ratios for this situation, in the Regge limit where $t_i,y_i,\vec{w}_i$ all scale toward zero, are
\bea
z \bz &\approx& \frac{4}{z_0^2}( 4t_0^2 + (\vec{w}_1 - \vec{w}_2)^2 + (y_1 - y_2)^2 ) \\
 z + \bz &\approx& 4t_0 (1 + \frac{1}{z_0^2})  - 2i(y_1 - y_2) (1 - \frac{1}{z_0^2}) \ .
\eea
On the $z$ and $\bz$ complex planes, the range of the smearing integrals lies within the shaded region of figure \ref{fig:zregion}. The cross ratios always stay in the right half-plane, Re $z,\bz > 0$. This is an important feature of this smearing procedure,  required in order to apply the chaos bound to derive causality constraints. It differs from the more natural-looking, real-momentum wavepackets used in \cite{Costa:2017twz, Kulaxizi:2017ixa}, which involve smearing into the bulk-point regime with Re $z,\bz < 0$.

\subsection{Shockwaves in the smeared OPE}
We assume $\phi$ is a light operator, with dimension $O(1)$. We will show that the smeared operators $W$ behave similarly to heavy operators, in the following sense: First, the stress tensor in the state $W(t_0)|0\rangle$ is exactly that of the shockwave \eqref{gravityT}, localized on a null plane, with energy 
\be\label{lighte}
E = \frac{i \Delta_\phi}{2t_0}\left(1  - \frac{d-1}{2\Delta_\phi}\right) \ .
\ee
(The shockwave is real for imaginary $t_0$).
Second, the leading contribution to the Regge OPE is
\be\label{smearedope}
\frac{\overline{W}(t_0) W(t_0) }{
\langle \overline{W}(t_0) W(t_0)\rangle} \approx 1 - i E z_0^2 \int_{-\infty}^\infty dv h_{vv}(u=0, v, z = z_0, \vec{x} = 0) \ .
\ee
On the right-hand side is the shockwave operator, just as in the heavy-scalar Regge OPE, \eqref{gravityope}.  This means that the smearing effectively projects out double-trace $[\phi\phi]$ operators, which would contribute at the same order if we just inserted local operators without smearing.  That is, the correlator $\langle \overline{W} W \phi' \phi'\rangle$, in the Regge limit, has contributions from $T$ and $[\phi'\phi']$, but no contributions from $[\phi\phi]$ at leading order. 

We will derive \eqref{smearedope} in $d=4$, from both gravity and CFT. The formula for energy in general $d$ \eqref{lighte} was obtained by computing $\langle \overline{W} \int du T_{uu} W\rangle$ and comparing to the prediction of  \eqref{smearedope}, and matches the energy in the OPE that we will find for $d=4$.

\subsection{Gravity derivation}
On the gravity side, \eqref{smearedope} is derived by smearing the vertex Witten diagram in \eqref{vertexfig}.  The smearing integrals depend only on the relative separations up to an overall factor, so we only need to smear one of the operators. The Gaussian factors also drop out since the integral is dominated near the origin of the $(t_0, y, \vec{w})$ coordinates. Let
\be
x_1 = (-u_{center},-v_{center}, \vec{0}), \qquad 
x_2 = (u,v,\vec{x})
\ee
with the coordinates given by \eqref{intco}. 
Performing the integrals, we find
\be
\frac{
\int dy d\vec{w} (t_0 + i y)^{-\Delta_\phi} \Pi_{\alpha\beta}(x_1, x_2; z', x')
}{
\int dy d\vec{w} (t_0 + i y)^{-\Delta_\phi} \langle \phi(x_1)\phi(x_2)\rangle
}
 =\frac{1}{2} h_{uu}^{Shock} \delta_\alpha^u \delta_\beta^u 
\ee
where $h_{uu}^{Shock}$ is the shockwave metric \eqref{swmetric}, with $E$ given by \eqref{lighte}. 

Stripping off the gravity propagator as we did in sections \ref{sss:witten}-\ref{sss:relation}, this is equivalent to the OPE formula \eqref{smearedope}.

\subsection{CFT derivation from conformal Regge theory}\label{ss:regsmear}
Now we will derive the same OPE formula \eqref{smearedope} in a large-$N$, holographic CFT, using conformal Regge theory \cite{Costa:2012cb}.\footnote{We thank S. Caron-Huot and A. Zhiboedov for discussions of conformal Regge theory and this calculation.} As usual, we assume a large gap in the spectrum of single trace higher spin primaries, so that the only contribution to the Regge amplitude comes from the graviton. Conformal Regge theory provides a way to compute 4-point functions that automatically includes double-trace as well as single-trace exchanges. In terms of 4-point correlators, the statement of \eqref{smearedope} is that the smeared 4-point function can be computed as follows: $(i)$ replace the wavepackets by local operators at the centers, and $(ii)$ drop double-trace $[\phi\phi]$ contributions.  We will demonstrate this by smearing the conformal Regge amplitude.

The derivation follows \cite{Costa:2012cb} and is very similar to \cite{Kulaxizi:2017ixa, Costa:2017twz} so we will be brief. The starting point in conformal Regge theory is the 4-point correlator written as a conformal partial-wave expansion
\begin{align}
G(z,\bz)\equiv\frac{\langle \phi\phi\psi\psi\rangle}{\langle \phi\phi\rangle \langle\psi\psi\rangle}=&\sum_{\Delta,J} C_{\phi\phi\O}C_{\psi\psi\O}g_{\Delta,J}(z,\bz)\notag\\
=&\sum_J\int d\nu~ b_J(\nu^2) F_{\nu,J}(z,\bz),
\label{blockexp}
\end{align}
where conformal partial waves $F_{\nu,J}$ are related to the conformal blocks by
\begin{align}
F_{\nu,J}(z,\bz)=&\kappa_{\nu,J}g_{h+i \nu,J}(z,\bz)+\kappa_{-\nu,J}g_{h-i \nu,J}(z,\bz),
\end{align}
with $h\equiv \frac{d}{2}$ and
\begin{align}
\kappa_{\nu,J}=	\frac{i \nu}{2\pi K_{h+i \nu,J}}
\end{align}
where $K_{h+i \nu,J}$ is given explicitly in \cite{Costa:2012cb}. We can evaluate the second line of \eqref{blockexp} by closing the contour in $\nu$ below the real axis since the partial waves vanish exponentially as $\nu\rightarrow -i\infty$. In order for the contour integral to evaluate to the conformal block expansion $b_J(\nu^2)$ must have poles \begin{align}
b_J(\nu^2)\sim \frac{C_{\phi\phi\O}C_{\psi\psi\O} K_{\Delta,J}}{\nu^2+(\Delta-h)^2} ~~\text{as}~~\nu^2\rightarrow(\Delta-h)^2.
\label{reggepole}
\end{align}
Note that since we are interested in the Regge limit of the correlator we will be concerned with the analytic continuation of the partial waves to the second sheet, taking $z$ around 1 with $\bz$ held fixed.

The next step is to analytically continue this expansion away from integer spins
\begin{align}
G=&\int_{\mathcal{C}} \frac{dJ}{2\pi i} \frac{1}{\sin(\pi J)}\int d\nu~ b_J(\nu^2) F_{\nu,J}(z,\bz),
\end{align}
where $\mathcal{C}$ is the contour enclosing the poles of the denominator corresponding to the integer spins. At large $N$, deforming this contour to enclose the Regge pole \eqref{reggepole} and separating out the identity contribution one obtains \cite{Costa:2012cb}
\begin{align}
G=1+\int d\nu~a(\nu)(z\bz)^{\frac{1-j(\nu)}{2}}\Omega_{i\nu}(\bz/z),
\end{align}
where
\begin{align}
a(\nu)=&\frac{\beta (\nu ) \gamma (\nu ) \gamma (-\nu ) \left(\pi ^{h-1} 2^{j(\nu )-1} e^{-\frac{i}{2} \pi  j(\nu )}\right)}{\sin \left(\frac{\pi  j(\nu )}{2} \right)},\notag\\
\beta(\nu)=&\frac{\pi  j'(\nu ) \text{K}_{\Delta (j(\nu )),j(\nu )}}{4 \nu } C_{\phi\phi j(\nu)}C_{\psi\psi j(\nu)} ,\notag\\
\gamma(\nu)=&\Gamma \left(\frac{2\Delta_{\phi}-h+i \nu +j(\nu )}{2}\right) \Gamma \left(\frac{2\Delta_{\psi}-h+i \nu +j(\nu )}{2}\right)\notag\\
\Omega_{i\nu}(x)=&\frac{i \nu  x^{\frac{1}{2}-\frac{i \nu }{2}}}{\pi ^2 (x-1)}-\frac{i \nu  x^{\frac{1}{2}+\frac{i \nu }{2}}}{\pi ^2 (x-1)}.
\end{align}
Assuming that the higher spin single-trace primary operators have scaling dimensions larger than $\Delta_{\text{gap}}$ and that the stress tensor is the lowest dimension spin-2 operator gives us an ansatz for the leading Regge trajectory
\begin{align}\label{stresstrajectory}
j(\nu)=2+\frac{1}{\Delta_{\text{gap}}^2}\left(\nu^2+\frac{\Delta_T^2}{4}\right)+\mathcal{O}(\Delta_{\text{gap}}^{-4}).
\end{align}
Using this ansatz we find
\begin{align}
\label{crtscalar}
G=1+C_{\phi\phi T}C_{\psi\psi T} \int d\nu \frac{90 \pi  \Gamma \left(\Delta _{\psi }-\frac{i \nu }{2}\right) \Gamma \left(\frac{i \nu }{2}+\Delta _{\psi }\right) \Gamma \left(\Delta _{\phi }-\frac{i \nu }{2}\right) \Gamma \left(\frac{i \nu }{2}+\Delta _{\phi }\right)}{\left(\nu ^2+4\right)\Gamma \left(\Delta _{\psi }-1\right) \Gamma \left(\Delta _{\psi }+1\right) \Gamma \left(\Delta _{\phi }-1\right) \Gamma \left(\Delta _{\phi }+1\right)}\frac{\Omega_{i\nu}(\bz/z)}{\sqrt{z\bz}}.
\end{align}
We can evaluate this expression by evaluating the contour integral for each term in $\Omega$ separately by closing the contour in the upper or the lower half plane accordingly. The two contributions are identical since $\Omega$ is symmetric under $\nu\rightarrow-\nu$. Therefore we will consider only one of the terms. Note that in the upper half plane the integrand has a pole at $\nu=2i$ corresponding to the stress-tensor contrubtion as well as poles at $\nu=2 i (n + \Delta_\phi)$ and $\nu=2 i (n + \Delta_\psi)$ for all positive integers $n$, corresponding to the exchange of the double trace operators.

We are now ready to apply the smearing procedure described in section \ref{smearing} to the correlator. To this end, we start by smearing the partial-wave\footnote{\label{relcoord}In this expression $y$ and $\vec{w}$ correspond to relative coordinates $y_1-y_2$ and $\vec{w}_1-\vec{w}_2$. Note that the integrand is independent of the absolute coordinates which leads to a volume integral divergence that is cancelled by the smeared 2-point function $\langle \overline{W}(t_0) W(t_0)\rangle$ in the denominator.}
\begin{align}
\tilde{\Omega}_{i\nu}(t_0,z_0)\equiv&\int dy d\vec{w}\frac{1}{\left( \frac{(t_0 + i y_1)(t_0 - i y_2)}{z_0^2}x_{12}^2\right)^{\Delta_\phi}}\frac{\Omega_{i\nu}(\bz/z)}{\sqrt{z\bz}}\notag\\
=&\int dy r dr \left(4 t_0^2+r^2+y^2\right)^{-\Delta_\phi}\left(\frac{i \nu  z^{-\frac{i \nu}{2}} \bz^{\frac{i \nu }{2}}}{\pi ^2 (z-\bz)}+\nu\leftrightarrow -\nu\right) .
\end{align}
Using \eqref{intco} and performing the radial and $y$ integrations we find
\begin{align}
\tilde{\Omega}_{i\nu}(t_0,z_0)=&\int dy\frac{\nu  z_0^{2 \left(\Delta _{\phi }-1\right)} \left(2 t_0+i y\right){}^{-\Delta _{\phi }+\frac{i \nu }{2}+1} \left(2 t_0 \left(z_0^2+1\right)-i y \left(z_0^2-1\right)\right){}^{-\Delta _{\phi }-\frac{i \nu }{2}}}{2 \pi ^2 \left(\nu +2 i \left(\Delta _{\phi }-1\right)\right)}\notag\\
&\times\, _2F_1\left(-\Delta _{\phi }+\frac{i \nu }{2}+1,\Delta _{\phi }+\frac{i \nu }{2};-\Delta _{\phi }+\frac{i \nu }{2}+2;\frac{2 t_0+i y}{2 t_0 \left(z_0^2+1\right)-i y \left(z_0^2-1\right)}\right)\notag\\&+(\nu\leftrightarrow -\nu)\notag\\
=& \left.-\frac{\pi  2^{5-4 \Delta _{\phi }} \Gamma \left(2 \Delta _{\phi }-2\right) t_0^{3-2 \Delta _{\phi }}}{\Gamma \left(\Delta _{\phi }-\frac{i \nu }{2}\right) \Gamma \left(\frac{i \nu }{2}+\Delta _{\phi }\right)}\frac{\Omega_{i\nu}(\bz/z)}{\sqrt{z\bz}} \right|_{\vec{w},y=0}+\nu\leftrightarrow -\nu\ .
\end{align}
This has two salient features. First, the smearing has the effect of replacing the partial wave $\Omega_{i\nu}$ by the partial wave evaluated at the wavepacket centers.  Second, the gamma functions produced by smearing cancel the gamma functions in \eqref{crtscalar}. Exactly the same thing occurred for real wavepackets in \cite{Kulaxizi:2017ixa,Costa:2017twz}. The Gamma-function poles correspond to double trace operators $[\phi\phi]$ in the OPE, so by smearing we have projected out these operators.  Translated into a statement about the Regge OPE, this is equivalent to \eqref{smearedope}.

The smearing can also be performed on the Regge amplitude $G_{Regge}(z,\bz)$, \ie after doing the $\nu$-integral. This illustrates more directly which regime of cross-ratios is important to project out double-traces. An example is worked out in appendix \ref{app:reggesmearing}.

\section{Smearing operators with spin}

Finally, we turn to the Regge OPE of spinning operators.  These also obey an equation like \eqref{smearedope}, with coefficients that depend on the polarizations. However, we won't actually derive the analogue of \eqref{smearedope}. Instead, we work with the four-point function, and just show by smearing the conformal Regge amplitude that double-trace operators are projected out. In other words, the smeared correlator is equal to the smeared stress tensor block in the Regge limit.

In \cite{Afkhami-Jeddi:2016ntf}, we showed that smearing the stress tensor block reproduces from CFT the graviton causality constraints of \cite{Camanho:2014apa}, such as the suppression of the Gauss-Bonnet term in 5d gravity, or in CFT language, the suppression of $\frac{a-c}{c}$ in large-$N$ CFTs with a large gap. In that paper, we assumed that smearing would project out double-trace operators and gave some physical motivation for this, but did not derive it. Our goal here is to close that gap by repeating the analysis of the previous section, but now for the spinning OPEs $TT \to T$ and $JJ \to T$, where $T$ is the stress tensor and $J$ is a conserved spin-1 current. This is rather technical but the basic idea is identical to the scalar case --- smearing puts the fields in the Witten diagram onto geodesics, and so the smeared correlator can be computed by just smearing the stress tensor conformal block.  The results, and the method, are also nearly identical to \cite{Costa:2017twz, Kulaxizi:2017ixa}, except we use the complex rather than real wavepackets. (It is not entirely clear why this gives the same final answer, since the smearing is over a different range of cross ratios. Presumably the two different smearings can be related by a contour deformation but we will not explore this.)

\subsection{Setup}

We restrict to $d=4$. Our conventions for the 3-point functions $\langle JJT\rangle$ and $\langle TTT\rangle$ follow the appendices of \cite{causality2}. $\langle JJT\rangle$ has two structures, with constant coefficients $n_s$ and $n_f$; $\langle TTT\rangle$ has three structures, with independent coefficients $n_s$, $n_f$, and $n_v$. The conformal collider bounds are $n_i \geq 0$.

It was shown in \cite{Costa:2011mg,Costa:2011dw} that correlation functions with spinning external operators can be expanded in terms of spinning conformal blocks which are related to derivatives of certain scalar conformal blocks 
\begin{align}
G^{\mu_1...\mu_l,\nu_1...\nu_l}\equiv& \frac{\langle \O^{\mu_1...\mu_l}(x_1)\O^{\nu_1...\nu_l}(x_2)\psi(x_3)\psi(x_4)\rangle}{(x_{12}^2)^{\Delta_{\O}+l}(x_{34}^2)^{\Delta_\psi}}\notag\\
=&\sum_{\Delta,J}C_{\psi\psi \O}\sum_k C_{\O_l\O_l \O}^{(k)}\hat{D}^{(k),{\mu_1...\mu_l,\nu_1...\nu_l}}g^{(k)}_{\Delta,J}(z,\bz),
\end{align}
where $\hat{D}^{(k),{\mu_1...\mu_l,\nu_1...\nu_l}}$ are differential operators acting on scalar conformal blocks $g^{(k)}_{\Delta,J}(z,\bz)$ with shifted scaling dimensions. Applying the methods of conformal Regge theory we will obtain the following expression for the Regge limit of the spinning correlator
\begin{align}
G^{\mu_1...\mu_l,\nu_1...\nu_l}=G_0^{\mu_1...\mu_l,\nu_1...\nu_l}+\int d\nu~\sum_k a^{(k)}(\nu)\hat{D}^{(k),{\mu_1...\mu_l,\nu_1...\nu_l}}(z\bz)^{\frac{1-j(\nu)}{2}}\Omega_{i\nu}(\bz/z),
\end{align}
where $G_0^{\mu_1...\mu_l,\nu_1...\nu_l}$ contains the identity contribution to the correlation function. In what follows we will use these methods to compute correlation functions involving spin 1 conserved currents, spin 2 stress tensors and scalars as  external operators.

\subsection{$JJ\psi\psi$} \label{JJsec}
Denote the normalized correlator by
\begin{align}
G^{\mu,\nu}\equiv& \frac{\langle J^{\mu}(x_1)J^{\n}(x_2)\psi(x_3)\psi(x_4)\rangle}{(x_{12}^2)^{\Delta_{J}+1}(x_{34}^2)^{\Delta_\psi}} \ .
\end{align}
Assuming the leading trajectory is given by \eqref{stresstrajectory}, the Regge amplitude (subtracting the identity) takes the form 
\begin{align}\label{jjlong1}
\epsilon_\mu\tilde{\epsilon}_\nu\delta G^{\mu,\nu}= \int d\nu~\sum_{k=1}^3 a^{(k)}(\nu)\epsilon_\mu\tilde{\epsilon}_\nu\hat{D}^{(k){\mu_,\nu}}\frac{\Omega_{i\nu}(\bz/z)}{\sqrt{z\bz}},
\end{align}
where the differential operators $\hat{D}^{(k){\mu_,\nu}}$ and coefficients $ a^{(k)}(\nu)$, as well as other details of this calculation, are given in appendix \ref{app:JJ}.

At this point we will assume that $\Delta_\psi$ is large so that we may neglect contribution from the $[\psi\psi]$ double-trace operators and perform the $\nu$ integral above by summing over the residues. Finally, we perform the smearing procedure described in section \eqref{smearing}. We choose coordinates for the operator insertions which are related to \eqref{intco} by a null inversion \eqref{fourshift},
\begin{align}
x_1=&(t_0,i y_1,i\vec{w}_1)\notag\\
x_2=&(-t_0,i y_2,i\vec{w}_2)\notag\\
x_3=&(1+z_0^2,1-z_0^2,\vec{0})\notag\\
x_4=&(-1-z_0^2,-1+z_0^2,\vec{0}).
\end{align}
In other words, we work directly in the shifted patch, where the wavepackets are simple complex momentum insertions.
We will also choose the following for the polarization vectors
\begin{align}
\epsilon=&\frac{1}{2 \sigma  z_0^2}\left(-1,1,-\lambda ,i \lambda \right)\notag\\
\tilde{\epsilon}=&\frac{1}{2 \sigma  z_0^2} \left(1, -1, \lambda, i\lambda\right).
\end{align}
We then perform the rescaling $t_0=\sigma t_0, y_i=\sigma y_i, \vec{w}_i=\sigma \vec{w}_i$ and take $\sigma\rightarrow 0$ limit. The resulting expression will depend only on relative coordinates $\vec{w}_{12}\equiv \vec{w}$ and $y_{12}\equiv y$ (see footnote \ref{relcoord}). We must now evaluate 
\begin{align}
\int dy d\vec{w}\epsilon_\mu \tilde{\epsilon}_\nu\frac{\delta G^{\mu,\nu}}{(x_{12}^2)^{\Delta_J+1}}.
\end{align}
Normalizing by the smeared 2-point function to regulate divergences we find
\begin{align}
&-\frac{15 i \pi ^3 \left(\lambda ^2-1\right) (4 n_f-n_s)}{16 (z_0^2 -1)^2 \sigma }-\frac{15 i \pi ^3 \left(\lambda ^2-1\right) (4 n_f-n_s)}{16 (z_0^2 -1)^3 \sigma }\notag\\
&-\frac{15 i \pi ^3 \left(2 \lambda ^2+1\right) (8 n_f+n_s)}{32 (z_0^2 -1) \sigma }-\frac{15 i \pi ^3 \left(2 \lambda ^2+1\right) (8 n_f+n_s)}{32 \sigma }
\end{align}
This is identical to the expression obtaining by ignoring, from the start, all of the double-trace poles in the conformal Regge amplitude. This is the main point: the smeared correlator is equal to the smeared stress tensor conformal block.

The chaos bound \cite{Maldacena:2015waa} implies that each power of $\lambda$ as $z_0 \to 1$ must be $i \times (positive)$ \cite{Afkhami-Jeddi:2016ntf}. Different polarizations give different signs, so this is impossible, unless
\begin{align}
4n_f-n_s=0.
\end{align}
In bulk language, this implies that the non-minimal coupling between the graviton and photon must be suppressed by the string scale. 

In terms of the Regge OPE, this implies a smeared OPE of $JJ$ similar to \eqref{smearedope}, but with the coefficient on the r.h.s built from polarizations, derivatives, and the $n_i$. It would be nice to work this out explicitly, for example by comparison to the bulk or to the 4-point Regge amplitude, but we have not done so. The combination $4n_f - n_s$ appears as the dominant coefficient in this OPE as $z_0 \to 1$.

\subsection{$TT\psi\psi$} \label{TTsec}
Now we turn to the case of external gravitons. Let
\begin{align}
G^{\mu\nu,\alpha\beta}\equiv& \frac{\langle T^{\mu\nu}(x_1)T^{\alpha\beta}(x_2)\psi(x_3)\psi(x_4)\rangle}{(x_{12}^2)^{\Delta_{T}+2}(x_{34}^2)^{\Delta_\psi}} \ .
\end{align}
Assuming the leading trajectory is given by \eqref{stresstrajectory}, subtracting the identity contribution and contracting with polarization vectors in the Regge limit we have
\begin{align}
\epsilon_\mu\epsilon_\nu\tilde{\epsilon}_\alpha\tilde{\epsilon}_\beta \delta G^{\mu\nu,\alpha\beta}=\int d\nu~\sum_{k=1}^{10} a^{(k)}(\nu)\epsilon_\mu\epsilon_\nu\tilde{\epsilon}_\alpha\tilde{\epsilon}_\beta\hat{D}^{(k){\mu\nu,\alpha\beta}}\frac{\Omega_{i\nu}(\bz/z)}{\sqrt{z\bz}},
\end{align}
where the differential operators $\hat{D}^{(k){\mu\nu,\alpha\beta}}$ and the coefficients $a^{(k)}(\nu)$ are too long to write down, but are available in Mathematica format from the authors. As above, we drop the heavy $[\psi\psi]$ operators, perform the $\nu$ integral by summing over residues, then do the smearing integrals. The result is
\begin{align}
&\frac{240 i \pi ^3 \left(\lambda ^4-4 \lambda ^2+1\right) (4 n_f-n_s-2 n_v)}{(z_0^2 -1)^4 \sigma }+\frac{120 i \pi ^3 \left(\lambda ^4-4 \lambda ^2+1\right) (4 n_f-n_s-2 n_v)}{(z_0^2 -1)^5 \sigma }\notag\\
&+\frac{20 i \pi ^3 \left(3 \lambda ^4 (14 n_f-n_s-22 n_v)+3 \lambda ^2 (-41 n_f+9 n_s+28 n_v)+27 n_f-8 n_s-6 n_v\right)}{(z_0^2 -1)^3 \sigma }\notag\\
&+\frac{20 i \pi ^3 \left(3 \left(\lambda ^4 (6 n_f+n_s-18 n_v)+\lambda ^2 (-9 n_f+n_s+12 n_v)+n_f\right)-2 n_s+6 n_v\right)}{(z_0^2 -1)^2 \sigma }\notag\\
&-\frac{10 i \pi ^3 \left(6 \left(\lambda ^4+\lambda ^2\right)+1\right) (6 n_f+n_s+12 n_v)}{(z_0^2 -1) \sigma }-\frac{10 i \pi ^3 \left(6 \left(\lambda ^4+\lambda ^2\right)+1\right) (6 n_f+n_s+12 n_v)}{\sigma }
\end{align}
Once again this agrees with the stress tensor term alone. In the leading term as $z_0\rightarrow 1$ the coefficients of powers of $\lambda$ come with opposite signs and hence cannot all be positive implying the following constraints on the coefficients
\begin{align}
4 n_f - n_s - 2 n_v=&0\notag\\
n_f - 2n_v =&0.
\end{align}
These are precisely the combinations corresponding to higher curvature couplings, ruled out from a gravity analysis in \cite{Camanho:2014apa}. Thus we have confirmed the argument sketched in \cite{Afkhami-Jeddi:2016ntf} that double trace operators can be ignored in this calculation.

\bigskip

\noindent \textbf{Acknowledgments}\\
It is a pleasure to thank Simon Caron-Huot, Daliang Li, and Sasha Zhiboedov for numerous discussions of double trace operators in the Regge limit, as well as Tom Faulkner, Diego Hofman, Jared Kaplan, Henry Maxfield, Sam McCandlish, and Arvin Moghaddam for essential conversations. The work of NAJ, TH, and AT is supported by DOE grant DE-SC0014123, and the work of SK is supported by NSF grant PHY-1316222.

\appendix

\section{Witten diagram calculations}\label{app:witten}

In this appendix, we derive some of the equations in section \ref{s:reggeope} by evaluating Witten diagrams in the Regge limit.  

\subsection{Feynman rules}

\subsubsection*{Scalar bulk-to-boundary propagator}
The scalar bulk-to-boundary propagator between a bulk point $(z,x)$ and a boundary point $x'$ in Euclidean AdS$_{d+1}$ is given by
\be
D(z,x;x')=c_\Delta \frac{z^\Delta}{(z^2+|\vec{x}-\vec{x}'|^2)^\Delta}\ ,
\ee
where
\be
c_\Delta=\frac{\Gamma(\Delta)}{\pi^{d/2}\Gamma(\Delta-d/2)}\ .
\ee
$\Delta=\frac{1}{2}(d+\sqrt{d+4m^2})$ is the scaling dimension of the boundary operator which is dual to a bulk scalar field of mass $m$.
\subsubsection*{Graviton bulk-to-bulk propagator}
Let us now write down the graviton bulk-to-bulk propagator between point $(z_1,y_1)$ and $(z_2,y_2)$ in Euclidean AdS$_{d+1}$. We first introduce the quantity 
\be
{\cal U}=-1+\frac{1}{2z_1 z_2}\left(z_1^2+z_2^2+(y_1-y_2)^2 \right)\ , \qquad \zeta=\frac{2z_1 z_2}{z_1^2+z_2^2+(y_1-y_2)^2}\ .
\ee
Let us also introduce 
\be
G({\cal U})=(8\pi G_N)\frac{\Gamma(d)\Gamma\left(\frac{d+1}{2} \right)\zeta^d}{2\pi^{(d+1)/2}\Gamma(d+1)} {}_2 F_1\left(\frac{d}{2}, \frac{1+d}{2},\frac{d}{2}+1,\zeta^2 \right)\ .
\ee
The bulk-to-bulk propagator is given by \cite{D'Hoker:1999pj}
\be
G_{\mu\nu\mu'\nu'}=\left(\partial_\mu\partial_{\mu'}{\cal U} \partial_\nu\partial_{\nu'}{\cal U}+\partial_\mu \partial_{\nu'}{\cal U} \partial_\nu\partial_{\mu'}{\cal U}\right) G({\cal U}) +g_{\mu\nu}g_{\mu'\nu'} H({\cal U})
\ee
where,
\be
H({\cal U})=-\frac{2(1+{\cal U})^2}{(d-1)} G({\cal U})+\frac{2(d-2)(1+{\cal U})}{(d-1)}\int_{{\cal U}}^\infty du' G(u')\ .
\ee

\subsubsection*{Graviton-scalar-scalar vertex}
The graviton-scalar-scalar vertex is given by the bulk stress tensor of the scalar field in terms of the scalar bulk-to-boundary propagators
\be\label{gss}
T^{bulk}_{\mu\nu}(D^\psi_1;D^\psi_2)=-{\cal N} \left(\partial_\mu D^\psi_1 \partial_\nu D^\psi_2+\partial_\nu D^\psi_1 \partial_\mu D^\psi_2 -(\partial_\alpha D^\psi_1)(\partial^\alpha D^\psi_2) g^{AdS}_{\mu\nu}- m^2 D^\psi_1 D^\psi_2 g^{AdS}_{\mu\nu}\right)\ .
\ee
where,
\be
D^\psi_1\equiv D^\psi(z,x;x_1)\ , \qquad D^\psi_2\equiv D^\psi(z,x;x_2)\ 
\ee
and 
\be\label{defN}
{\cal N}=\frac{1}{2 c_{\Delta_\psi} (2\Delta_\psi-d)}\ .
\ee
Note that all the derivatives in equation (\ref{gss}) are taken with respect to the bulk point $(z,x)$.

\subsection{Vertex diagram}

Now consider the scalar-scalar-graviton vertex diagram,  shown in \eqref{vertexfig}. We set $d=4$, but the result easily generalizes. $\psi$ is a heavy operator with $c_T \gg \Delta_\psi \gg 1$. In general, this diagram is given by 
\be
\Pi_{\alpha' \beta'}(x_1,x_2;z',x')=i \int d^{d}x dz \sqrt{-g^{AdS}} G^{\mu\nu}_{\alpha' \beta'}(z,x; z',x')T^{bulk}_{\mu\nu}(D^\psi(z,x;x_1);D^\psi(z,x;x_2))\ ,
\ee
where, $G^{\mu\nu}_{\alpha' \beta'}(z,x; z',x')$ is the graviton bulk-to-bulk propagator and $D^\psi(z,x;x')$ is the scalar bulk-to-boundary propagator. $T^{bulk}_{\mu\nu}$ is the bulk stress tensor of the dual scalar field in terms of the scalar bulk-to-boundary propagator (\ref{gss}).  $\Pi_{\alpha' \beta'}(x_1,x_2;z',x')$ can easily be computed by using the results of \cite{DHoker:1999mqo} (and also \cite{Fitzpatrick:2011hh}). For, $d=4$, following \cite{DHoker:1999mqo}, we can write down 
\be\label{pi}
\Pi_{\alpha' \beta'}(x_1,x_2;z',x')=-\frac{ \Delta_\psi}{4\pi^2}\frac{8\pi G_N}{x_{12}^{2\Delta_\psi}}\frac{1}{z'^2}\left(\frac{1}{3}\eta_{\alpha' \beta'}-J_{\alpha' z'}(X'-X_1)J_{\beta' z'}(X'-X_1) \right)f(t)+\cdots\ ,
\ee
where $X_1=(z=0,x_1),\ X'=(z,x')$ and dots represent gauge dependent terms which will not contribute to the final answer. The inversion tensor $J_{\alpha \beta}(X'-X_1)$ is given by
\be
J_{\alpha \beta}(X'-X_1)=\eta_{\alpha \beta}-\frac{2(X'-X_1)_\alpha (X'-X_1)_\beta}{z^2+|x'-x_1|^2}\ .
\ee
with indices raised and lowered by $\eta_{\alpha\beta}$.
The function $f(t)$ is given by \cite{DHoker:1999mqo, Fitzpatrick:2011hh}
\be
f(t)=\frac{t(1-t^{\Delta_\psi-1})}{(1-t)}\ , \qquad t=\frac{z'^2 x_{12}^2}{(z'^2+(x'-x_1)^2) (z'^2+(x'-x_2)^2)}\ .
\ee
Note that equation (\ref{pi}) is not symmetric with respect to $x_1$ and $x_2$. This asymmetry, as noted in  \cite{Fitzpatrick:2011hh}, is a consequence of dropping the gauge dependent terms. In an actual correlator, there will be an integral over $(z',x')$ which will make the final answer symmetric under $x_1 \leftrightarrow x_2$.

So far we have not assume anything about $\Delta_\psi$. Let us now take the limit $\Delta_\psi\rightarrow \infty$. In this limit we can approximate 
\be\label{fapprox}
f(t)\approx \frac{t}{1-t}\ .
\ee
In the Euclidean signature, $t\le 1$ and hence the term $t^{\Delta_\psi}$ can be ignored.\footnote{Here we compute the Euclidean answer, then analytically continue. This is valid as long as we do not go so far into the Regge regime as to compete with the large-$\Delta_\psi$ limit. The same final formula can be obtained by taking $u,v$ imaginary, as in the shockwave state discussed in section \ref{s:dictionary}, and evaluating the diagram by a saddlepoint approximation.}

Let us now choose $x_1=(u,v,\vec{0})$ and $x_2=-x_1$ with $u>0, v<0$. In the Regge limit (\ref{reggedef}), using (\ref{pi}) along with (\ref{fapprox}), we obtain
\be\label{reggeshock}
\Pi_{\alpha' \beta'}(x_1,x_2;z',x')=i\frac{(8\pi G_N)\Delta_\psi}{16\pi}\frac{1}{(-4uv)^{\Delta_\psi}}\frac{u}{z' \sqrt{-u v}}\frac{  (r-1)^3 }{   r (r+1) } \delta_{\alpha' }^v\delta_{\beta' }^v\delta(v')+\O\left(u^0, \frac{1}{\Delta_\psi} \right)\ ,
\ee
where (recall that $x'=(u',v',\vec{x}')$),
\be
r=\sqrt{\frac{\vec{x}'^2+\left(z'-\sqrt{-u v }\right)^2}{\vec{x}'^2+\left(z'+\sqrt{-u v }\right)^2}}\ .
\ee
This is exactly the planar shockwave solution \eqref{swmetric} in AdS$_5$, with an additional factor of $i$, once we identify $z_0=\sqrt{-uv}$. Thus we have derived \eqref{vertexpi}.

To rewrite this as the null geodesic integral \eqref{regge2}, note that the bulk-to-bulk graviton propagator $G_{uu\alpha' \beta'}(u,v=0, \vec{x}=0, z=z_0; z',x')$ can be written as $P_{uu \alpha'\beta'}G(u=0,v, \vec{x}=0, z=z_0; z',x')$, where $P_{uu \alpha'\beta'}$ is independent of $u$ and $G(u=0,v, \vec{x}=0, z=z_0; z',x')$ is the scalar bulk-to-bulk propagator. Therefore,
\begin{align}
\int du G_{uu\alpha' \beta'}(u,v=0, \vec{x}=0, z=z_0; z',x')&=P_{uu \alpha'\beta'} \int du G(u,v=0, \vec{x}=0, z=z_0; z',x')\nonumber\\
&\equiv P_{uu \alpha'\beta'} g(z',v',\vec{x}')\ ,
\end{align}
where, in $(d+1)-$dimensions $g(z',v',\vec{x}')$ satisfies the differential equation
\be
\left(\partial_{z'}^2-(d-1)\frac{\partial_{z'}}{z'} +\partial_{\vec{x'}^2} \right)g(z',v',\vec{x}')=-2 i \left(z_0\right)^{d-1}\delta(v')\delta(\vec{x}')\delta(z'-z_0)\ .
\ee
$g(z',v',\vec{x}')$ satisfies exactly the same differential equation as a planar shockwave in AdS$_{d+1}$  (see for example \cite{Camanho:2014apa}) along with the same boundary condition and hence 
\be
g(z',v',\vec{x}')=2 i\delta(v') \frac{z' z_0(4\pi)^{\frac{1-d}{2}}\Gamma(\frac{d+1}{2})}{d(d-1)}\left(\frac{\rho^2}{1-\rho^2} \right)^{1-d} {~}_2 F_1 \left(d-1, \frac{d+1}{2},d+1,-\frac{1-\rho^2}{\rho^2}\right)\ ,
\ee
where,
\be
\rho=\sqrt{\frac{(z'-z_0)^2+\vec{x}'^2}{(z'+z_0)^2+\vec{x}'^2}}\ .
\ee
Therefore, with $d=4$,
\be\label{diagram1}
\int du G_{uu\alpha' \beta'}(u,v=0, \vec{x}=0, z=z_0; z',x')=-\frac{i G_N}{ z' z_0}\frac{  (\rho-1)^3 }{  \rho (\rho+1) } \delta_{\alpha' }^v\delta_{\beta' }^v\delta(v')
\ee
which allows us to rewrite (\ref{reggeshock}) as
\be\label{regge2app}
\Pi_{\alpha' \beta'}(x_1,x_2;z',x')=-\frac{\Delta_\psi u}{2} \int_{-\infty}^{\infty} du'' G_{uu\alpha' \beta'}(u'',v=0, \vec{x}=0, z=\sqrt{-uv}; z',x')\ .
\ee
This is \eqref{regge2} in the main text.

\subsection{4-point functions}

Now consider the four-point function $\langle \psi (u,v) \O(x_3) \O(x_4) \psi(-u,-v)\rangle$ in the Regge limit, where $\O$ is an arbitrary operator with or without spin. This four-point function can  be computed from the bulk side using Witten diagrams, involving the vertex function computed above. The exchange diagram gives
\begin{align}
\langle \psi (u,v) \O(x_3)\O(x_4)  \psi(-u,-v)  \rangle& =\langle \O(x_3) \O(x_4)\rangle \langle \psi (x_1) \psi(x_2)\rangle\\
&+ 2i\int d^{4}x' dz' \sqrt{-g^{AdS}}T_{\O}^{\alpha' \beta'}(z,x';x_3,x_4)\Pi_{\alpha' \beta'}+\cdots\ , \nonumber
\end{align}
where $T_{\O}^{\alpha' \beta'}$ is the bulk stress tensor of the field dual to the operator $\O$. Using (\ref{reggeshock}), this implies
\be\label{shockoo}
\frac{\langle \psi (u,v) \O(x_3)\O(x_4) \psi(-u,-v)  \rangle_{\text Regge}}{ \langle \psi (x_1) \psi(x_2)\rangle}=\langle \O(x_3) \O(x_4)\rangle_{shock}\ ,
\ee 
where $\langle \O(x_3) \O(x_4)\rangle_{shock}$ is the two-point function computed in the imaginary shockwave \eqref{imshock}.

\section{Check of the stress-tensor block}\label{app:doblock}

In this appendix, we use the Regge OPE (\ref{regge_hkll}) to reproduce the scalar conformal block for stress tensor exchange in the Regge limit in $d=4$. Let us consider the correlator
\be\label{appcor}
\langle \psi (u,v) \phi (x=-1) \phi (x=1) \psi (-u,-v) \rangle\ .
\ee
In the Regge limit (\ref{reggedef}), the contribution from the stress tensor according to (\ref{regge_hkll}) is 
\begin{align}\label{block}
g_T(z,\bz)&=\frac{2 \lambda_T }{\pi^2 v}\int_{t'^2+x'^2+\vec{x}'^2<-uv}dt' dx' d^2 \vec{x}' \nonumber\\
& \times \int_{-\infty}^{\infty} du' \langle \phi (x=-1) T_{uu}\left(\frac{u'}{2}+t',-\frac{u'}{2}+i x',i \vec{x}'\right) \phi (x=1)\rangle \ ,
\end{align}
where in the Regge limit $z=4/u,\ \bz=-4v$. The $u$-integral is done first and then the remaining integral. The $u$-integral is subtle, as discussed in section \ref{sss:ordering}: It is defined in such a way that the $u$-contour circles one pole. First, let us write
\be
u=-\eta \sigma\ ,\qquad v=\frac{1}{\sigma}
\ee
where, the Regge limit is obtained by taking $\sigma \rightarrow 0$. Note that the conformal cross-ratios in the Regge limit are
\be\label{zzbar}
\bz=4\eta \sigma\ , \qquad z=4\sigma\ .
\ee
We now perform the $u$-integral. If we do the $u$-integral along the real line, then the $u$-contour does not enclose any poles and hence the integral vanishes. Instead, to obtain the ordering \eqref{appcor}, the $u$-integral must be done along the following contour:
\be\label{ucontour}
\begin{gathered}\includegraphics[width=0.6\textwidth]{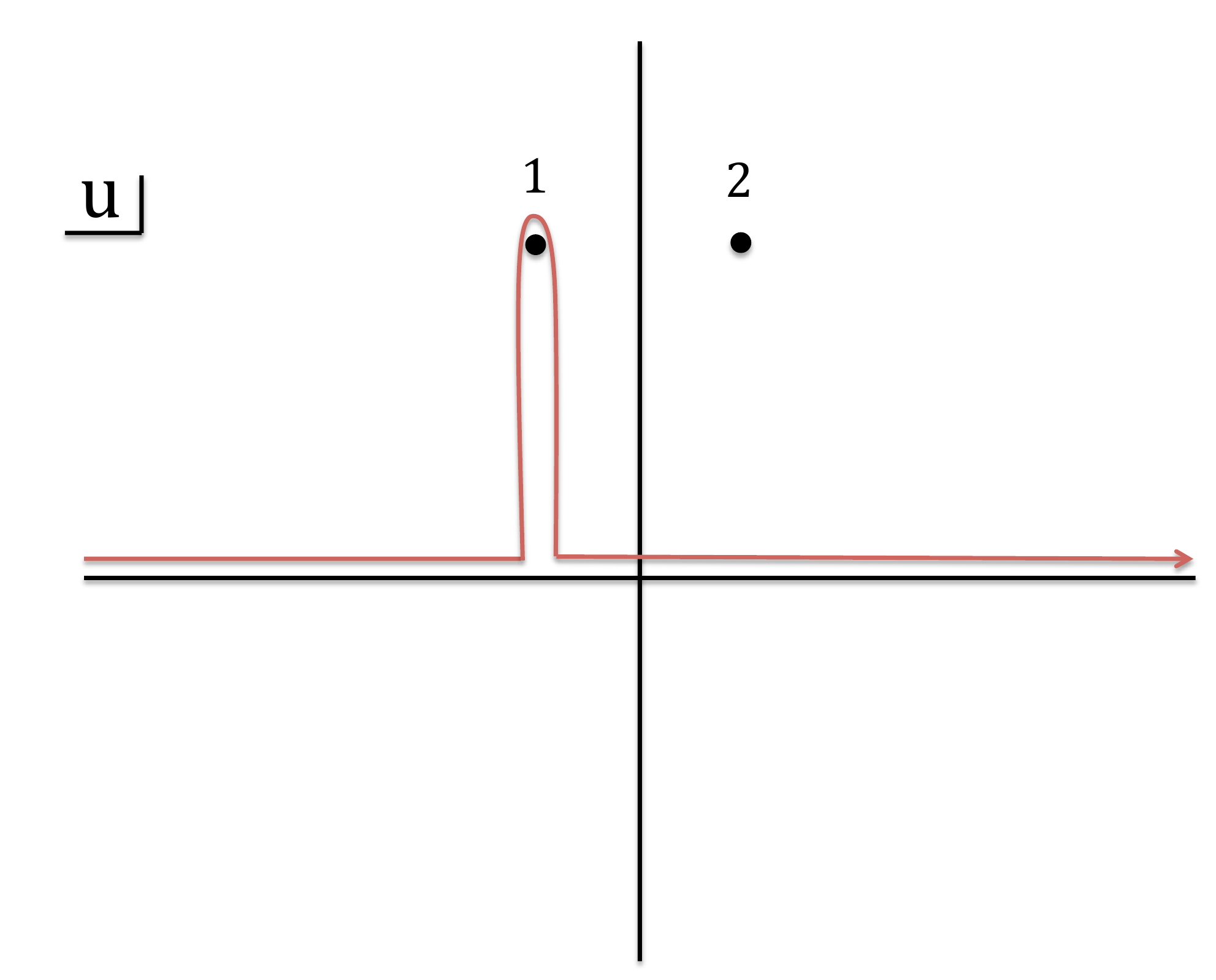}\end{gathered}
\ee
where the poles $1$ and $2$ are due to the operators $ \phi (x=1)$ and $ \phi (x=-1)$ respectively. This means that the $u$ integral can be evaluated by a residue, with the result
\be
g_T=\frac{10 \Delta _{\psi } \Delta _{\phi } }{\pi^6 c_T v}\int_{t'^2+x'^2+\vec{x}'^2<-uv}dt' dx' d^2 \vec{x}' \frac{(t'+i x'-1)^2 (t'+i x'+1)^2}{\left(t'^2+2 i t' x'-x'^2+\vec{x}'^2-1\right)^3}\ .
\ee
Now, one can perform the $\vec{x}'$-integrals and then $t'$ and $x'$ integrals, yielding 
\be\label{finalblock}
g_T(z,\bz)= i\left(\frac{40 \Delta_\psi \Delta_\O}{c_T \pi^3}\right)\frac{ \bz }{z (z-\bz)}\ .
\ee
Note that we could have also used \eqref{tblock} because of the way the contours are defined.

Let's compare \eqref{finalblock} to the known conformal block, as computed by Dolan and Osborn \cite{Dolan:2000ut}. In 4d, the conformal block for stress tensor exchange, including the OPE coefficient set by the Ward identity, is
\be\label{doblockt}
g_T^{full}(z,\bz) = -\left(\frac{4\Delta_\psi\Delta_O}{9c_T \pi^4}\right) \frac{z^4 \bz \, {}_2F_1(3,3,6,z) - \bz^4 z \, {}_2F_1(3,3,6,\bz)}{4(z-\bz)}
\ee
The hypergeometric function is $_2F_1(3,3,6,z) = \frac{90}{z^4}(z-2) - \frac{30}{z^5}(z^2-6z+6)\log(1-z)$. To reach the Regge limit, we first take $\log (1-z) \to \log(1-z) + 2\pi i $, then $z,\bz \to 0$. The leading term is exactly \eqref{finalblock}.

\subsection{Time delay from the Dolan-Osborn block}\label{app:dotimedelay}
 In section \ref{sss:timedelaycft} we derived the gravitational time delay $\Delta v$ using the Regge OPE. Here, for comparison, we repeat the CFT calculation in $d=4$ using the full Dolan-Osborn conformal block, quoted in \eqref{doblockt}. The cross-ratios in the kinematics \eqref{geokina}, \eqref{phikina}, with $\vec{x}_\phi = 0$ and $\delta \ll 1$, are
 \be\label{ffcross}
 z = -\frac{4i \delta}{u_\phi}, \qquad \bz = \frac{4iv_\phi \delta}{z_0^2} \ .
 \ee
 Plugging these values into \eqref{finalblock} gives exactly the gravity result without winding, on the first line of \eqref{g4final}. This is the case where we reach the Regge regime by sending $z$ around 1. If we instead send $\bz$ around 1, or equivalently exchange $z \leftrightarrow \bz$ in \eqref{ffcross} and apply \eqref{finalblock}, then the result matches the second line of \eqref{g4final}.

 \section{Derivation of the general Lightcone and Regge block}\label{app:generalregge}
In this appendix, we find a general formula for the OPE block of (scalar)$\times$(scalar) $\to$ (anything)  in the Regge limit, assuming $x_1$ and $x_2$ are timelike separated. The result is analytically continued to spacelike separation in \eqref{genreg}.  Using the shadow operator formalism, the OPE block is written as an integral over a causal diamond, in the future of $x_1$ and past of $x_2$. In lightcone or Regge limit, the diamond will shrink to a line or a slab respectively, and the OPE formula simplifies.

Following \cite{Czech:2016xec,deBoer:2016pqk}, we write the OPE as
\begin{align}
&\frac{\psi(x_1) \psi(x_1) } {\langle  \psi(x_2) \psi(x_1) \rangle}= \mathcal{N}_X \int_{D(x_1,x_2) } d^d\zeta \left(\frac{(x_1-\zeta)^2 (x_2-\zeta)^2}{x_{21}^2}\right)^{\frac{\Delta-d}{2}}  \hat{t}_{12}^{\mu_1 } \hat{t}_{12}^{\mu_2 } \cdots \hat{t}_{12}^{\mu_l } \; X_{\mu_1  \mu_2 \cdots \mu_\ell}
 \end{align}
where $\mathcal{N}_X$ is a normalization constant to be determined below, and
 \begin{align}
&t_{12}^{\mu} = \hat{t}_{12}^{\mu} |t_{12}| = \frac{(x_2 -\zeta)^{\mu} }{(x_2 - \zeta)^2} - \frac{(x_1 - \zeta)^{\mu} }{(x_1 -\zeta)^2}  \ .
\end{align}
The kinematics are the same as the main text,
\begin{align}
x_2 = (u,v, \vec{0})\ , \qquad
x_1 = (-u ,-v ,\vec{0})\ , \qquad
\zeta= ( \tilde{u},\tilde{v}, \vec{x})\ .
\end{align}
However, here we assume both $u$ and  $v$ are positive so the diamond $D(x_1,x_2)$ is real. The diamond is the region in spacetime defined by
\begin{align}
 \frac{\vec{x}^2} {\tilde{u}+u } - v  \le \tilde{v} \le v  -\frac{\vec{x}^2}{u -\tilde{u}}\ ,  \qquad \vec{x}^2 \le u v \left(1-\frac{\tilde{u}^2}{u^2}\right)\ , \qquad-u \le \tilde{u} \le u\ .
\end{align}
In both the Regge and lightcone limits, $v \to 0$, so $\tilde{v} \to 0 $ as well. Hence, we can set $\tilde{v}=0$ in the argument of $X$ and integrate the expression over $v$.  For the same reason, we keep only $\mu = u$ in $t^{\mu}_{12}$. 
In either limit, integration over $\tilde{v}$ yields:
\begin{align}\label{eq:regge1}
&\frac{\psi(x_2) \psi(x_1)|_X}{\langle \psi(x_2) \psi(x_1) \rangle}  \sim 
 \frac{u^{\ell-1}}{(u v )^ {\frac{\Delta+ \ell -d}{2}}} \int du d^{d-2}\vec{x}   \frac{\left[ u v \left(1-\frac{\tilde{u}^2}{u^2} \right) - \vec{x}^2 \right]^{\Delta -d+1}}{\left(1-\frac{\tilde{u}^2}{u^2}\right)^{\frac{\Delta-d-\ell}{2}+1}} X_{uu\cdots u}(u,0,\vec{x})\ .
\end{align}
In the lightcone limit, $uv \to 0$, one can also integrate over the transverse part in \ref{eq:regge1} and we recover the OPE as a line integral appearing in \cite{Hartman:2016lgu}:
\begin{align}
\frac{\psi(x_2) \psi(x_1)|^{L.C}_X}{\langle \psi(x_2) \psi(x_1) \rangle}  \sim u^{\ell-1} (u v )^{\frac{\Delta -\ell}{2}}  \int_{-u}^{u} d\tilde{u}  (1-\tilde{u}^2/u^2)^{\frac{\Delta+\ell}{2} - 1} X_{uu\cdots u}(\tilde{u},0,\vec{0})\ .
\end{align}
Taking the Regge limit of \eqref{eq:regge1}, $v\to 0,  u v= \text{fixed}$, we find:
\begin{align}
\frac{\psi(x_2) \psi(x_1)|^{\text{Regge}}_X}{\langle \psi(x_2) \psi(x_1) \rangle}  \sim    \frac{(u v)^{\frac{d-\Delta-\ell}{2}}}{u ^{1-\ell}} \int_{-\infty}^{+\infty} d\tilde{u} \int_{\vec{x}^2 \le uv} d^{d-2}\vec{x} \left( u v - \vec{x}^2\right)^{\Delta-d+1} X_{uu\cdots u} (\tilde{u},0,\vec{x}) \ .
\end{align}
Finally by taking $uv \to 0$ after the Regge limit and comparing to the lightcone result in \cite{Hartman:2016lgu}, we find the constant of proportionality. The final answer including factors is:
\begin{align}
& \frac{\psi(x_2)\psi(x_1)|^{\text{Regge}}_X}{\langle \psi(x_2) \psi(x_1) \rangle } = (-1)^{\frac{\Delta-l}{2}} \pi^{\frac{1 -d}{2}} 2^{\Delta}  \frac{\Gamma(\frac{\Delta+\ell+1}{2})}{\Gamma(\frac{\Delta+\ell}{2})} \frac{\Gamma(\Delta -d/2+1)}{\Gamma(\Delta-d+2)} \\
& \quad \qquad \frac{C_{\psi \psi X}}{C_X} \frac{(u v)^{\frac{ d-\Delta-\ell}{2}}}{u^{1-\ell}} \int_{-\infty}^{+\infty} d\tilde{u} \int_{\vec{x}^2 \le u v}d^{d-2}\vec{x} ( u v -\vec{x}^2)^{\Delta-d+1} X_{uu\cdots u}(\tilde{u},0,\vec{x})\ ,\notag
\end{align}
where $C_{\psi\psi X}$ is the OPE coefficient and $C_X$ is the normalization of the two-point function $\langle XX\rangle$.


\section{Smearing the Regge amplitude}
\label{app:reggesmearing}

In this appendix, we give a simple example that illustrates how smearing projects out double-trace operators in the Regge amplitude. This gives another perspective on the results of section \ref{ss:regsmear}.

Consider the correlator $\langle \phi \phi \psi \psi\rangle$, where $\phi$ and $\psi$ are scalar primaries with $\Delta_\phi = 2$ and $\Delta_\psi \gg 1$. The Regge amplitude corresponding to graviton exchange in the bulk, obtained by doing the $\nu$ integral in \eqref{crtscalar}, is
\be
G -1 \sim \frac{z \bz}{(z + \bz)^3} \ .
\ee
(We will not keep track of the overall factor). Note that, unlike the stress tensor conformal block, this is regular at $z = \bz$; the apparent singularity from stress tensor exchange is smoothed out by the double trace contributions. Nonetheless we will find singular behavior at $z_{center} = \bz_{center}$ after smearing.  

The smearing integral, with the kinematics from section \ref{ss:regsmear}, is then 
\bea
\delta G &\sim& \int_0^\infty rdr \int_{-\infty}^{\infty}dy 
\frac{1}{\left( \frac{(t_0 + i y_1)(t_0 - i y_2)}{z_0^2}x_{12}^2\right)^{\Delta_\phi}}\frac{z \bz}{(z + \bz)^3}\\
&=& \int_0^\infty r dr \int_{-\infty}^{\infty}dy\frac{z_0^2}{2 \left(r^2+4 t_0^2+y^2\right) \left(2 t_0 \left(z_0^2+1\right)-i y \left(z_0^2-1\right)\right){}^3}\\
&=& \frac{\pi }{16 t_0 } \frac{z_{center}}{\bz_{center}(z_{center} - \bz_{center})}  \ .
\eea

Smearing just the stress tensor block gives the same result, so the double-trace operators have indeed been projected out.

\section{Regge amplitude for $\langle JJ \psi \psi\rangle$ }\label{app:JJ}
In this appendix we give some details of the calculation in section \ref{JJsec}.
The differential operators $\hat{D}^{(1){\mu_,\nu}}$ are given explicitly by
\fontsize{8}{8}\selectfont
\begin{align}
\epsilon_\mu\epsilon_\nu\hat{D}^{(1){\mu_,\nu}}=&u^2 V_{124} V_{213} \partial^2_u-v^2 (V_{123}-V_{124}) (V_{213}-V_{214}) \partial^2_v\notag\\
&+u  \left(V_{124} V_{213}-\frac{1}{2} H_{12}\right)\partial_u-v (V_{123}-V_{124}) (V_{213}-V_{214}) \partial_v\notag\\
&-u v (V_{123} V_{213}+V_{124} (V_{214}-2 V_{213})) \partial_u\partial_v\ ,\notag\\
\epsilon_\mu\epsilon_\nu\hat{D}^{(2){\mu_,\nu}}=&H_{12}\ ,\notag\\
\epsilon_\mu\epsilon_\nu\hat{D}^{(3){\mu_,\nu}}=&u^2 v V_{123} V_{214} \partial^2_u+u  \left(v V_{123} V_{214}-\frac{1}{2} H_{12}\right)\partial_u\notag\\
&-u v (V_{123} (V_{213}-2 v V_{214})+V_{124} V_{214})\partial_u\partial_v\notag\\
&+ (v V_{123}-V_{124}) (v V_{214}-V_{213})\partial_v\notag\\
&+v  (v V_{123}-V_{124}) (v V_{214}-V_{213})\partial^2_v\ ,
\end{align}
\normalsize
with $u=z\bz,v=(1-z)(1-\bz)$ and conformal structures defined as
\begin{align}
H_{ij}&=-2 \left(-(x_{ij}\cdot \epsilon_i)(x_{ji}\cdot \epsilon_j)-\frac{1}{2} (\epsilon_i\cdot \epsilon_j) (x_{ij}\cdot x_{ij})\right)\ ,\notag\\
V_{ijk}&=-\frac{2 \left(\frac{1}{2} (x_{ij}\cdot x_{ij}) (x_{ki}\cdot \epsilon_i)-\frac{1}{2} (x_{ik}\cdot x_{ik}) (x_{ji}\cdot \epsilon_i)\right)}{x_{jk}\cdot x_{jk}}\ .
\end{align}
The coefficients $a^{(k)}(\nu)$ introduced in \ref{JJsec} are given by
\begin{align}
a^{(1)}(\nu)=&\frac{10 \pi  \nu  n_f \Gamma \left(4-\frac{i \nu }{2}\right) \Gamma \left(\frac{i \nu }{2}+4\right) \Gamma \left(\Delta_\psi -\frac{i \nu }{2}\right) \Gamma \left(\Delta_\psi +\frac{i \nu }{2}\right)}{\left(\nu ^2+4\right) \Gamma (\Delta_\psi -1) \Gamma (\Delta_\psi +1)}\ ,\notag\\
a^{(2)}(\nu)=&\frac{5 \pi  \nu  \Gamma \left(4-\frac{i \nu }{2}\right) \Gamma \left(\frac{i \nu }{2}+4\right) (12 n_f+n_s) \Gamma \left(\Delta_\psi -\frac{i \nu }{2}\right) \Gamma \left(\Delta_\psi +\frac{i \nu }{2}\right)}{2 \left(\nu ^2+4\right) \Gamma (\Delta_\psi -1) \Gamma (\Delta_\psi +1)}\ ,\notag\\
a^{(3)}(\nu)=&-\frac{5 \pi  \nu  \Gamma \left(4-\frac{i \nu }{2}\right) \Gamma \left(\frac{i \nu }{2}+4\right) (4 n_f+n_s) \Gamma \left(\Delta_\psi -\frac{i \nu }{2}\right) \Gamma \left(\Delta_\psi +\frac{i \nu }{2}\right)}{2 \left(\nu ^2+4\right) \Gamma (\Delta_\psi -1) \Gamma (\Delta_\psi +1)}\ .
\end{align}
The full Regge amplitude in \eqref{jjlong1} is
\fontsize{7}{7}\selectfont
\begin{align}
&\epsilon_\mu \tilde{\epsilon}_\nu\delta G^{\mu,\nu}=-2\int d\nu\frac{5 \pi  \nu  \Gamma \left(4-\frac{i \nu }{2}\right) \Gamma \left(\frac{i \nu }{2}+4\right) z^{-\frac{i \nu}{2}} \bz^{\frac{i \nu }{2}} \Gamma \left(\Delta_\psi -\frac{i \nu }{2}\right) \Gamma \left(\Delta_\psi +\frac{i \nu }{2}\right)}{8 \left(\nu ^2+4\right) \Gamma (\Delta_\psi -1) \Gamma (\Delta_\psi +1) (z-\bz)^5}\notag\\
&4 n_f (12 H_{12} z^2+\nu ^2 V_{124} V_{213} z^2-4 i \nu  V_{124} V_{213} z^2-4 V_{124} V_{213} z^2-24 \bz H_{12} z-2 \bz \nu ^2 V_{124} V_{213} z-8 \bz V_{124} V_{213} z+8 i \nu  V_{124} V_{213} z\notag\\&
+8 V_{124} V_{213} z+12 \bz^2 H_{12}-4 \bz^2 V_{124} V_{213}+\bz^2 \nu ^2 V_{124} V_{213}+8 \bz V_{124} V_{213}+4 i \bz^2 \nu  V_{124} V_{213}-8 i \bz \nu  V_{124} V_{213}\notag\\&
-(z-1) (\bz-1) ((z-\bz)^2 \nu ^2-4 i (z-\bz) (z+\bz-2) \nu -4 (z+\bz-2) (z+\bz)) V_{123} V_{214}) (z-\bz)^2+n_s (((-i \nu +\bz (2 i \nu +2)+4) H_{12}\notag\\&
+(\bz-1) (V_{123} ((4-2 \nu  (-3 i+\nu )) V_{213}+(5 \nu ^2+4 i (\bz-5) \nu +8 (\bz-2)) V_{214})-2 (\nu  (-3 i+\nu )-2) V_{124} V_{214})) z^4+2 ((\nu ^2+12) V_{123} V_{214} \bz^3\notag\\&
+((-3 i \nu -1) H_{12}+(\nu  (3 i+\nu )+10) V_{123} V_{213}+((\nu  (3 i+\nu )+10) V_{124}-(\nu  (2 i+3 \nu )+34) V_{123}) V_{214}) \bz^2+(i (10 i+\nu ) H_{12}\notag\\&
+2 (-2 i+\nu ) ((-4 i+\nu ) V_{123}-(-i+\nu ) V_{124}) V_{213}-2 (-4 i+\nu ) ((-4 i+\nu ) V_{123}-(-2 i+\nu ) V_{124}) V_{214}) \bz+(-i+\nu ) (V_{123} (2 (-5 i+2 \nu ) V_{214}\notag\\&
-3 (-2 i+\nu ) V_{213})+V_{124} (2 (-i+\nu ) V_{213}-3 (-2 i+\nu ) V_{214}))) z^3+2 (2 (2-i \nu ) V_{123} V_{214} \bz^4+((3 i \nu -1) H_{12}+(-5 i+\nu ) (2 i+\nu ) V_{123} V_{213}\notag\\&
+(((2 i-3 \nu ) \nu -34) V_{123}+(-5 i+\nu ) (2 i+\nu ) V_{124}) V_{214}) \bz^3+(16 H_{12}+2 (2 (\nu ^2+4) V_{124}-3 (\nu ^2+6) V_{123}) V_{213}+((9 \nu ^2+64) V_{123}\notag\\&
-6 (\nu ^2+6) V_{124}) V_{214}) \bz^2+(3 (\nu  (3 i+\nu )+10) V_{123}-2 (\nu  (6 i+\nu )+11) V_{124}) V_{213} \bz+(3 (\nu  (3 i+\nu )+10) V_{124}-2 (\nu  (3 i+2 \nu )+19) V_{123}) V_{214} \bz\notag\\&
+2 (-2 i+\nu ) (-i+\nu ) (V_{123}-V_{124}) (V_{213}-V_{214})) z^2+\bz (((2-2 i \nu ) H_{12}-2 (i+\nu ) (2 i+\nu ) V_{124} V_{214}+V_{123} ((\nu  (24 i+5 \nu )-24) V_{214}\notag\\&
-2 (i+\nu ) (2 i+\nu ) V_{213})) \bz^3+2 ((-i \nu -10) H_{12}+2 (2 i+\nu ) V_{124} ((4 i+\nu ) V_{214}-(i+\nu ) V_{213})+2 (4 i+\nu ) V_{123} ((2 i+\nu ) V_{213}\notag\\&
-(4 i+\nu ) V_{214})) \bz^2+2 (V_{124} (3 (2 i+\nu ) (-5 i+\nu ) V_{214}-2 (\nu  (-6 i+\nu )+11) V_{213})+V_{123} (3 (2 i+\nu ) (-5 i+\nu ) V_{213}-2 (\nu  (-3 i+2 \nu )+19) V_{214})) \bz\notag\\&
-\bz^4 (2 i+\nu )^2 V_{123} V_{214}+8 (\nu ^2+4) (V_{123}-V_{124}) (V_{214}-V_{213})) z-z^5 (\bz-1) (-2 i+\nu )^2 V_{123} V_{214}+\bz^2 ((2 i+\nu )^2 V_{123} V_{214} \bz^3+((i \nu +4) H_{12}\notag\\&
+2 (i+\nu ) (2 i+\nu ) V_{123} V_{213}+((16-5 \nu  (4 i+\nu )) V_{123}+2 (i+\nu ) (2 i+\nu ) V_{124}) V_{214}) \bz^2+2 (i+\nu ) (V_{123} (2 (5 i+2 \nu ) V_{214}-3 (2 i+\nu ) V_{213})\notag\\&
+V_{124} (2 (i+\nu ) V_{213}-3 (2 i+\nu ) V_{214})) \bz+4 (i+\nu ) (2 i+\nu ) (V_{123}-V_{124}) (V_{213}-V_{214}))).
\end{align}
\normalsize
Evaluating the $\nu$ integral by residues, with $\Delta_\psi \gg 1$, gives
\fontsize{7}{7}\selectfont
\begin{align}
&\epsilon_\mu \tilde{\epsilon}_\nu\delta G^{\mu,\nu}=\frac{2}{(z+\bz)^9}120 i \pi  z \bz (\frac{1}{2} H_{12} (z+\bz) (24 n_f (z+\bz) (z^4+8 z^3 \bz+29 z^2 \bz^2+8 z \bz^3+\bz^4)+n_s (z^5 (3 \bz+1)+z^4 \bz (32 \bz+5)\notag\\&
+z^3 \bz^2 (163 \bz-5)+z^2 \bz^3 (32 \bz-5)+z \bz^4 (3 \bz+5)+\bz^5))-V_{124} V_{213} (8 n_f (2 z^6+3 z^5 (8 \bz-1)+z^4 \bz (138 \bz-35)-z^3 \bz^2 (188 \bz+195)\notag\\&
+3 z^2 \bz^3 (46 \bz-65)+z \bz^4 (24 \bz-35)+\bz^5 (2 \bz-3))-n_s (3 z^5 (4 \bz-3)+z^4 (\bz (168 \bz-145)+12)+3 z^3 \bz (384 \bz^2-375 \bz+56)\notag\\&
+3 z^2 \bz^2 (\bz (56 \bz-375)+384)+z \bz^3 (\bz (12 \bz-145)+168)+3 \bz^4 (4-3 \bz)))+(z-1) (\bz-1) V_{123} V_{214} (8 n_f (2 z^6+3 z^5 (8 \bz-1)\notag\\&
+z^4 \bz (138 \bz-35)-z^3 \bz^2 (188 \bz+195)+3 z^2 \bz^3 (46 \bz-65)+z \bz^4 (24 \bz-35)+\bz^5 (2 \bz-3))+n_s (4 z^6+15 z^5 (4 \bz-1)+z^4 (\bz (444 \bz-215)+12)\notag\\&
+z^3 \bz (\bz (776 \bz-1515)+168)+3 z^2 \bz^2 (\bz (148 \bz-505)+384)+z \bz^3 (5 \bz (12 \bz-43)+168)+\bz^4 (\bz (4 \bz-15)+12)))\notag\\&
+6 n_s (z-1) (\bz-1) V_{123} V_{213} (z+\bz-2) (z^4+14 z^3 \bz+96 z^2 \bz^2+14 z \bz^3+\bz^4)\notag\\&
+6 n_s (z-1) (\bz-1) V_{124} V_{214} (z+\bz-2) (z^4+14 z^3 \bz+96 z^2 \bz^2+14 z \bz^3+\bz^4))\ .
\end{align}
\normalsize

\end{spacing}


\begin{thebibliography}{99}%

\bibitem{Aichelburg:1970dh} 
  P.~C.~Aichelburg and R.~U.~Sexl,
  ``On the Gravitational field of a massless particle,''
  Gen.\ Rel.\ Grav.\  {\bf 2}, 303 (1971).
  doi:10.1007/BF00758149

\bibitem{Hotta:1992qy} 
  M.~Hotta and M.~Tanaka,
  ``Shock wave geometry with nonvanishing cosmological constant,''
  Class.\ Quant.\ Grav.\  {\bf 10}, 307 (1993).
  doi:10.1088/0264-9381/10/2/012

\bibitem{Horowitz:1999gf} 
  G.~T.~Horowitz and N.~Itzhaki,
  ``Black holes, shock waves, and causality in the AdS / CFT correspondence,''
  JHEP {\bf 9902}, 010 (1999)
  [hep-th/9901012].

\bibitem{Cornalba:2006xk} 
  L.~Cornalba, M.~S.~Costa, J.~Penedones and R.~Schiappa,
  ``Eikonal Approximation in AdS/CFT: From Shock Waves to Four-Point Functions,''
  JHEP {\bf 0708}, 019 (2007)
  doi:10.1088/1126-6708/2007/08/019
  [hep-th/0611122].

\bibitem{Cornalba:2006xm} 
  L.~Cornalba, M.~S.~Costa, J.~Penedones and R.~Schiappa,
  ``Eikonal Approximation in AdS/CFT: Conformal Partial Waves and Finite N Four-Point Functions,''
  Nucl.\ Phys.\ B {\bf 767}, 327 (2007)
  doi:10.1016/j.nuclphysb.2007.01.007
  [hep-th/0611123].

\bibitem{Cornalba:2007zb} 
  L.~Cornalba, M.~S.~Costa and J.~Penedones,
  ``Eikonal approximation in AdS/CFT: Resumming the gravitational loop expansion,''
  JHEP {\bf 0709}, 037 (2007)
  doi:10.1088/1126-6708/2007/09/037
  [arXiv:0707.0120 [hep-th]].

\bibitem{Hofman:2008ar} 
  D.~M.~Hofman and J.~Maldacena,
  ``Conformal collider physics: Energy and charge correlations,''
  JHEP {\bf 0805}, 012 (2008)
  doi:10.1088/1126-6708/2008/05/012
  [arXiv:0803.1467 [hep-th]].

\bibitem{Hofman:2009ug} 
  D.~M.~Hofman,
  ``Higher Derivative Gravity, Causality and Positivity of Energy in a UV complete QFT,''
  Nucl.\ Phys.\ B {\bf 823}, 174 (2009)
  doi:10.1016/j.nuclphysb.2009.08.001
  [arXiv:0907.1625 [hep-th]].



\bibitem{Nozaki:2013wia} 
  M.~Nozaki, T.~Numasawa and T.~Takayanagi,
  ``Holographic Local Quenches and Entanglement Density,''
  JHEP {\bf 1305}, 080 (2013)
  doi:10.1007/JHEP05(2013)080
  [arXiv:1302.5703 [hep-th]].
  


\bibitem{Shenker:2013pqa} 
  S.~H.~Shenker and D.~Stanford,
  ``Black holes and the butterfly effect,''
  JHEP {\bf 1403}, 067 (2014)
  doi:10.1007/JHEP03(2014)067
  [arXiv:1306.0622 [hep-th]].

\bibitem{Asplund:2014coa} 
  C.~T.~Asplund, A.~Bernamonti, F.~Galli and T.~Hartman,
  ``Holographic Entanglement Entropy from 2d CFT: Heavy States and Local Quenches,''
  JHEP {\bf 1502}, 171 (2015)
  [arXiv:1410.1392 [hep-th]].

\bibitem{Camanho:2014apa} 
  X.~O.~Camanho, J.~D.~Edelstein, J.~Maldacena and A.~Zhiboedov,
  ``Causality Constraints on Corrections to the Graviton Three-Point Coupling,''
  JHEP {\bf 1602}, 020 (2016)
  doi:10.1007/JHEP02(2016)020
  [arXiv:1407.5597 [hep-th]].

\bibitem{Kulaxizi:2017ixa} 
  M.~Kulaxizi, A.~Parnachev and A.~Zhiboedov,
  ``Bulk Phase Shift, CFT Regge Limit and Einstein Gravity,''
  arXiv:1705.02934 [hep-th].


\bibitem{Costa:2017twz} 
  M.~S.~Costa, T.~Hansen and J.~Penedones,
  ``Bounds for OPE coefficients on the Regge trajectory,''
  arXiv:1707.07689 [hep-th].

\bibitem{causality1} 
  T.~Hartman, S.~Jain and S.~Kundu,
  ``Causality Constraints in Conformal Field Theory,''
  JHEP {\bf 1605}, 099 (2016)
  doi:10.1007/JHEP05(2016)099
  [arXiv:1509.00014 [hep-th]].
  
\bibitem{causality2} 
  T.~Hartman, S.~Jain and S.~Kundu,
  ``A New Spin on Causality Constraints,''
  arXiv:1601.07904 [hep-th].

\bibitem{Hartman:2016lgu} 
  T.~Hartman, S.~Kundu and A.~Tajdini,
  ``Averaged Null Energy Condition from Causality,''
  arXiv:1610.05308 [hep-th].

\bibitem{Afkhami-Jeddi:2016ntf} 
  N.~Afkhami-Jeddi, T.~Hartman, S.~Kundu and A.~Tajdini,
  ``Einstein gravity 3-point functions from conformal field theory,''
  arXiv:1610.09378 [hep-th].

\bibitem{Roberts:2014isa} 
  D.~A.~Roberts, D.~Stanford and L.~Susskind,
  ``Localized shocks,''
  JHEP {\bf 1503}, 051 (2015)
  doi:10.1007/JHEP03(2015)051
  [arXiv:1409.8180 [hep-th]].
  






  
\bibitem{Caputa:2014vaa} 
  P.~Caputa, M.~Nozaki and T.~Takayanagi,
  ``Entanglement of local operators in large-N conformal field theories,''
  PTEP {\bf 2014}, no. 9, 093B06 (2014)
  [arXiv:1405.5946 [hep-th]].
  






  
  
\bibitem{Brower:2006ea} 
  R.~C.~Brower, J.~Polchinski, M.~J.~Strassler and C.~I.~Tan,
  ``The Pomeron and gauge/string duality,''
  JHEP {\bf 0712}, 005 (2007)
  doi:10.1088/1126-6708/2007/12/005
  [hep-th/0603115].


\bibitem{Cornalba:2007fs} 
  L.~Cornalba,
  ``Eikonal methods in AdS/CFT: Regge theory and multi-reggeon exchange,''
  arXiv:0710.5480 [hep-th].


\bibitem{Cornalba:2008qf} 
  L.~Cornalba, M.~S.~Costa and J.~Penedones,
  ``Eikonal Methods in AdS/CFT: BFKL Pomeron at Weak Coupling,''
  JHEP {\bf 0806}, 048 (2008)
  doi:10.1088/1126-6708/2008/06/048
  [arXiv:0801.3002 [hep-th]].

\bibitem{Costa:2012cb} 
  M.~S.~Costa, V.~Goncalves and J.~Penedones,
  ``Conformal Regge theory,''
  JHEP {\bf 1212}, 091 (2012)
  doi:10.1007/JHEP12(2012)091
  [arXiv:1209.4355 [hep-th]].

\bibitem{Heemskerk:2009pn} 
  I.~Heemskerk, J.~Penedones, J.~Polchinski and J.~Sully,
  ``Holography from Conformal Field Theory,''
  JHEP {\bf 0910}, 079 (2009)
  doi:10.1088/1126-6708/2009/10/079
  [arXiv:0907.0151 [hep-th]].
  
  
  





\bibitem{Komargodski:2012ek} 
  Z.~Komargodski and A.~Zhiboedov,
  ``Convexity and Liberation at Large Spin,''
  JHEP {\bf 1311}, 140 (2013)
  doi:10.1007/JHEP11(2013)140
  [arXiv:1212.4103 [hep-th]].

\bibitem{Fitzpatrick:2012yx} 
  A.~L.~Fitzpatrick, J.~Kaplan, D.~Poland and D.~Simmons-Duffin,
  ``The Analytic Bootstrap and AdS Superhorizon Locality,''
  JHEP {\bf 1312}, 004 (2013)
  doi:10.1007/JHEP12(2013)004
  [arXiv:1212.3616 [hep-th]].
  
\bibitem{Alday:2016htq} 
  L.~F.~Alday and A.~Bissi,
  ``Unitarity and positivity constraints for CFT at large central charge,''
  arXiv:1606.09593 [hep-th].

\bibitem{Balakrishnan:2017bjg} 
  S.~Balakrishnan, T.~Faulkner, Z.~U.~Khandker and H.~Wang,
  ``A General Proof of the Quantum Null Energy Condition,''
  arXiv:1706.09432 [hep-th].

\bibitem{Kabat:2012hp} 
  D.~Kabat, G.~Lifschytz, S.~Roy and D.~Sarkar,
  ``Holographic representation of bulk fields with spin in AdS/CFT,''
  Phys.\ Rev.\ D {\bf 86}, 026004 (2012)
  doi:10.1103/PhysRevD.86.026004, 10.1103/PhysRevD.86.029901
  [arXiv:1204.0126 [hep-th]].


\bibitem{Czech:2016xec} 
  B.~Czech, L.~Lamprou, S.~McCandlish, B.~Mosk and J.~Sully,
  ``A Stereoscopic Look into the Bulk,''
  JHEP {\bf 1607}, 129 (2016)
  doi:10.1007/JHEP07(2016)129
  [arXiv:1604.03110 [hep-th]].

\bibitem{deBoer:2016pqk} 
  J.~de Boer, F.~M.~Haehl, M.~P.~Heller and R.~C.~Myers,
  ``Entanglement, holography and causal diamonds,''
  JHEP {\bf 1608}, 162 (2016)
  doi:10.1007/JHEP08(2016)162
  [arXiv:1606.03307 [hep-th]].

\bibitem{Maldacena:2015waa} 
  J.~Maldacena, S.~H.~Shenker and D.~Stanford,
  ``A bound on chaos,''
  JHEP {\bf 1608}, 106 (2016)
  doi:10.1007/JHEP08(2016)106
  [arXiv:1503.01409 [hep-th]].

\bibitem{Caron-Huot:2017vep} 
  S.~Caron-Huot,
  ``Analyticity in Spin in Conformal Theories,''
  arXiv:1703.00278 [hep-th].

\bibitem{Engelhardt:2016aoo} 
  N.~Engelhardt and S.~Fischetti,
  ``The Gravity Dual of Boundary Causality,''
  Class.\ Quant.\ Grav.\  {\bf 33}, no. 17, 175004 (2016)
  doi:10.1088/0264-9381/33/17/175004
  [arXiv:1604.03944 [hep-th]].

\bibitem{Gao:2000ga} 
  S.~Gao and R.~M.~Wald,
  ``Theorems on gravitational time delay and related issues,''
  Class.\ Quant.\ Grav.\  {\bf 17}, 4999 (2000)
  doi:10.1088/0264-9381/17/24/305
  [gr-qc/0007021].


\bibitem{Hofman:2016awc} 
  D.~M.~Hofman, D.~Li, D.~Meltzer, D.~Poland and F.~Rejon-Barrera,
  ``A Proof of the Conformal Collider Bounds,''
  JHEP {\bf 1606}, 111 (2016)
  doi:10.1007/JHEP06(2016)111
  [arXiv:1603.03771 [hep-th]].


\bibitem{Kelly:2014mra} 
  W.~R.~Kelly and A.~C.~Wall,
  ``Holographic proof of the averaged null energy condition,''
  Phys.\ Rev.\ D {\bf 90}, no. 10, 106003 (2014)
  Erratum: [Phys.\ Rev.\ D {\bf 91}, no. 6, 069902 (2015)]
  doi:10.1103/PhysRevD.90.106003, 10.1103/PhysRevD.91.069902
  [arXiv:1408.3566 [gr-qc]].

\bibitem{Faulkner:2016mzt} 
  T.~Faulkner, R.~G.~Leigh, O.~Parrikar and H.~Wang,
  ``Modular Hamiltonians for Deformed Half-Spaces and the Averaged Null Energy Condition,''
  JHEP {\bf 1609}, 038 (2016)
  doi:10.1007/JHEP09(2016)038
  [arXiv:1605.08072 [hep-th]].

   \bibitem{dhss}
  S.~de Haro, S.~N.~Solodukhin and K.~Skenderis,
  ``Holographic reconstruction of spacetime and renormalization in the  AdS/CFT
  correspondence,''
  Commun.\ Math.\ Phys.\  {\bf 217} (2001) 595
  [arXiv:hep-th/0002230].

 \bibitem{skenderis}
  K.~Skenderis,
  ``Asymptotically anti-de Sitter spacetimes and their stress energy  tensor,''
  Int.\ J.\ Mod.\ Phys.\  A {\bf 16}, 740 (2001)
  [arXiv:hep-th/0010138].

  \bibitem{skenderis2}
  K.~Skenderis,
  ``Lecture notes on holographic renormalization,''
  Class.\ Quant.\ Grav.\  {\bf 19} (2002) 5849
  [arXiv:hep-th/0209067].

\bibitem{Osborn:1993cr} 
  H.~Osborn and A.~C.~Petkou,
  ``Implications of conformal invariance in field theories for general dimensions,''
  Annals Phys.\  {\bf 231}, 311 (1994)
  [hep-th/9307010].

\bibitem{Anous:2017tza} 
  T.~Anous, T.~Hartman, A.~Rovai and J.~Sonner,
  ``From Conformal Blocks to Path Integrals in the Vaidya Geometry,''
  arXiv:1706.02668 [hep-th].
  
  
\bibitem{maxfieldtalk}
H. Maxfield, talk at QIQG3 workshop, Vancouver, August 2017.



\bibitem{Hijano:2015zsa} 
  E.~Hijano, P.~Kraus, E.~Perlmutter and R.~Snively,
  ``Witten Diagrams Revisited: The AdS Geometry of Conformal Blocks,''
  JHEP {\bf 1601}, 146 (2016)
  doi:10.1007/JHEP01(2016)146
  [arXiv:1508.00501 [hep-th]].
  
\bibitem{Penrose:1993ud} 
  R.~Penrose, R.~D.~Sorkin and E.~Woolgar,
  ``A Positive mass theorem based on the focusing and retardation of null geodesics,''
  gr-qc/9301015.
  
\bibitem{Woolgar:1993zp} 
  E.~Woolgar,
  ``Positive energy for asymptotically anti-de Sitter spaces,''
  gr-qc/9404020.
  
\bibitem{Kleban:2001nh} 
  M.~Kleban, J.~McGreevy and S.~D.~Thomas,
  ``Implications of bulk causality for holography in AdS,''
  JHEP {\bf 0403}, 006 (2004)
  doi:10.1088/1126-6708/2004/03/006
  [hep-th/0112229].

\bibitem{Brigante:2007nu} 
  M.~Brigante, H.~Liu, R.~C.~Myers, S.~Shenker and S.~Yaida,
  ``Viscosity Bound Violation in Higher Derivative Gravity,''
  Phys.\ Rev.\ D {\bf 77}, 126006 (2008)
  doi:10.1103/PhysRevD.77.126006
  [arXiv:0712.0805 [hep-th]].



\bibitem{Brigante:2008gz} 
  M.~Brigante, H.~Liu, R.~C.~Myers, S.~Shenker and S.~Yaida,
  ``The Viscosity Bound and Causality Violation,''
  Phys.\ Rev.\ Lett.\  {\bf 100}, 191601 (2008)
  doi:10.1103/PhysRevLett.100.191601
  [arXiv:0802.3318 [hep-th]].



\bibitem{Papallo:2015rna} 
  G.~Papallo and H.~S.~Reall,
  ``Graviton time delay and a speed limit for small black holes in Einstein-Gauss-Bonnet theory,''
  JHEP {\bf 1511}, 109 (2015)
  doi:10.1007/JHEP11(2015)109
  [arXiv:1508.05303 [gr-qc]].

\bibitem{Casini:2010bf} 
  H.~Casini,
  ``Wedge reflection positivity,''
  J.\ Phys.\ A {\bf 44}, 435202 (2011)
  doi:10.1088/1751-8113/44/43/435202
  [arXiv:1009.3832 [hep-th]].

\bibitem{Costa:2011mg} 
  M.~S.~Costa, J.~Penedones, D.~Poland and S.~Rychkov,
  ``Spinning Conformal Correlators,''
  JHEP {\bf 1111}, 071 (2011)
  doi:10.1007/JHEP11(2011)071
  [arXiv:1107.3554 [hep-th]].
  
\bibitem{Costa:2011dw} 
  M.~S.~Costa, J.~Penedones, D.~Poland and S.~Rychkov,
  ``Spinning Conformal Blocks,''
  JHEP {\bf 1111}, 154 (2011)
  doi:10.1007/JHEP11(2011)154
  [arXiv:1109.6321 [hep-th]].




\bibitem{D'Hoker:1999pj} 
  E.~D'Hoker, D.~Z.~Freedman, S.~D.~Mathur, A.~Matusis and L.~Rastelli,
  ``Graviton exchange and complete four point functions in the AdS / CFT correspondence,''
  Nucl.\ Phys.\ B {\bf 562}, 353 (1999)
  doi:10.1016/S0550-3213(99)00525-8
  [hep-th/9903196].


\bibitem{DHoker:1999mqo} 
  E.~D'Hoker, D.~Z.~Freedman and L.~Rastelli,
  ``AdS / CFT four point functions: How to succeed at z integrals without really trying,''
  Nucl.\ Phys.\ B {\bf 562}, 395 (1999)
  doi:10.1016/S0550-3213(99)00526-X
  [hep-th/9905049].
 
  
\bibitem{Fitzpatrick:2011hh} 
  A.~L.~Fitzpatrick and D.~Shih,
  ``Anomalous Dimensions of Non-Chiral Operators from AdS/CFT,''
  JHEP {\bf 1110}, 113 (2011)
  doi:10.1007/JHEP10(2011)113
  [arXiv:1104.5013 [hep-th]].
  
\bibitem{Dolan:2000ut} 
  F.~A.~Dolan and H.~Osborn,
  ``Conformal four point functions and the operator product expansion,''
  Nucl.\ Phys.\ B {\bf 599}, 459 (2001)
  doi:10.1016/S0550-3213(01)00013-X
  [hep-th/0011040].

\end{thebibliography}
\end{document}